\documentclass[%
 reprint,
 amsmath,amssymb,
 aps,
]{revtex4-2}

\usepackage{graphicx}
\usepackage{dcolumn}
\usepackage{bm}

\usepackage{subfigure}

\begin{document}

\preprint{APS/123-QED}

\title{Mixing of X and Y  states  from QCD Sum Rules analysis }

\author{Ze-Sheng Chen$^1$,~Zhuo-Ran Huang$^2$,~Hong-Ying Jin$^3$,~T.G. Steele$^4$,~and~Zhu-Feng Zhang$^5$}
\affiliation{
 Institute of Modern Physics, Department of Physics, Zhejiang University, Hangzhou, 310027,China$^{1,3}$\\Institute of High Energy Physics, Chinese Academy of Sciences, Beijing, 100049, China$^2$\\Department of Physics and Engineering Physics, University of Saskatchewan, Saskatoon, Saskatchewan, S7N 5E2, Canada$^4$\\Department of Physics, Ningbo University, Ningbo, 315211, China$^5$
}

\date{\today}

\begin{abstract}
We study $\bar{Q}Q\bar{q}q$  and $\bar{Q}qQ\bar{q}$ states as mixed states in QCD sum rules. By calculating the two-point correlation functions of pure states of their corresponding currents, we review the mass and coupling constant predictions of $J^{PC}=1^{++}$, $1^{--}$, $1^{-+}$  states. 
By calculating the two-point mixed correlation functions of $\bar{Q}Q\bar{q}q$ and $\bar{Q}qQ\bar{q}$  currents, and we estimate the mass and coupling constants of the corresponding ``physical state" that couples to both $\bar{Q}Q\bar{q}q$ and $\bar{Q}qQ\bar{q}$ currents. Our results suggest that $1^{++}$ states are more likely mixing from $\bar{Q}Q\bar{q}q$ and $\bar{Q}qQ\bar{q}$ components, while for $1^{--}$ and $1^{-+}$ states, there is less mixing between $\bar{Q}Q\bar{q}q$ and $\bar{Q}qQ\bar{q}$.  Our results suggest the $Y$ series of states have more complicated components.
\end{abstract}

\maketitle


\section{\label{sec:level1}Introduction}

\noindent
In 2003, Belle observed a new state, known as the $X(3872)$, which definitely contained a charm-anticharm pair and cannot be explained by ordinary quark-antiquark model~\cite{Brambilla:2019esw}. Since then, more new hadrons containing heavy quarks have been found and studied in numerous experiments~\cite{Zyla:2020zbs}. These hadrons are known as XYZ states, containing a heavy quark-antiquark pair and at least a light quark-antiquark pair, are naturally exotics~\cite{Albuquerque:2018jkn}. Many structures have been proposed to describe XYZ states including molecular, tetraquark and hybrid components ~\cite{Chen:2010ze,Lu:2016cwr,Chen:2013pya}. Like the studies of other mesons with exotic quantum numbers, the convincing explanations of observed XYZ states remains an open question in phenomenological particle physics. Recently, a study of $X(3872)$ by LHCb argued that the compact component should be required in X states\cite{LHCb:2020sey}, and this result is more likely to support the tetraquark model to XYZ states, but not exclude the molecular model to all exotic states. In this paper we focus on  states in two simple $\bar{Q}qQ\bar{q}$ and $\bar{Q}Q\bar{q}q$ combinations to study XYZ states ($\bar{Q}$ represents a heavy quark $c$ or $b$ quark while $q$ represents a light quark $u,d$ or $s$ quark). These two  forms have been extensively studied in previous papers~\cite{Albuquerque:2011ix,Finazzo:2011he}. However, $\bar{Q}qQ\bar{q}$ and $\bar{Q}Q\bar{q}q$ states are difficult to  distinguish straightforwardly from decay modes of XYZ states because XYZ states are usually observed to have both  $\bar{Q}Q+\bar{q}q$ like decay modes and $\bar{Q}q+Q\bar{q}$ like decay modes. Many scenarios were studied to distinguish $\bar{Q}qQ\bar{q}$ and $\bar{Q}Q\bar{q}q$  states qualitatively~\cite{Close:1995eu,Suzuki:2005ha,Matheus:2006xi,Thomas:2008ja,Liu:2008fh}. Furthermore, the physical states are usually mixtures of different  structures,  which makes the problem more complicated. In a previous study we have developed a method to estimate mixing strength of different currents from a QCD sum rule (QCDSR) approach~\cite{Chen:2019buw,Harnett:2008cw,Palameta:2017ols,Palameta:2018yce}, and we use the same technique to study the mixing of $\bar{Q}qQ\bar{q}$ and $\bar{Q}Q\bar{q}q$ XY states.

In this paper, we study three kinds of vector states with different quantum numbers $J^{PC}=1^{++}$, $1^{--}$, $1^{-+}$.
These states have long been considered to be $\bar{Q}qQ\bar{q}$ or $\bar{Q}Q\bar{q}q$ molecular states in different studies~\cite{Wang:2013daa,Wang:2014gwa,Gong:2016hlt,Albuquerque:2015kia,Cui:2013xla,Chen:2013omd,Chen:2015ata}. However, since many of them have abundant  decay modes, mixing scenarios should be taken into account. Besides, it is generally believed that there is a large background of two free mesons spectrum in the two points correlation function of  four-quark currents. To avoid such large uncertainty,
we especially estimate the mixing strength of $\bar{Q}qQ\bar{q}$ and $\bar{Q}Q\bar{q}q$ currents, to investigate the corresponding physical states. The calculations will show us whether the physical states prefer to be $\bar{Q}qQ\bar{q}$ or $\bar{Q}Q\bar{q}q$ molecular states, or whether they are strongly mixed states.

As mentioned above, for the 1$^{++}$ channel, $X(3872)$
has been extensively studied for a wide variety of structures~\cite{Liu:2008tn,Dong:2009yp}.
 In molecular state schemes, $X(3872)$ has been usually considered as $ D^* D$ molecular state~\cite{Matheus:2006xi,Thomas:2008ja,Liu:2008fh}.
However, although the pure $ D^* D$ molecular state was predicted to have mass close to $X(3872)$, it had a too large decay width to agree with the experimental results~\cite{Matheus:2009vq,Jackson:1979eu}. On the other hand, $J/\psi \rho$ or $J/\psi \omega$ state also have similar mass since their sum of masses of two constituent parts are close to $X(3872)$. Hence the mixing of these two molecular states is naturally possible. Besides, in recent observation to $X(3872)$ by LHCb\cite{LHCb:2020sey}, the compact component is required. Hence we will consider another state $\bar{c}c$ in the mixing, which have been studied in \cite{Matheus:2009vq}.

For the 1$^{--}$ channel, many 1$^{--}$ states are found in the range 4200-4700 MeV, permitting  abundant possible pure or mixed molecular states. Some 1$^{--}$ states have very similar mass like $Y(4220)/Y(4260)$ and $Y(4360)/Y(4390)$~\cite{Wang:2016wwe,Wang:2018rfw}. Hence it is interesting and meaningful to investigate the possible mixing of molecular state, which has not been previously studied.

For 1$^{-+}$ sector, no confirmed heavy hadrons with 1$^{-+}$ quantum numbers have been observed. Some potential candidates include $X(3940), X(4160), X(4350)$~\cite{Zyla:2020zbs}. The constructions of 1$^{-+}$ molecular states in $\bar{Q}qQ\bar{q}$ and $\bar{Q}Q\bar{q}q$ scenarios are possible. As outlined below, we calculate the mass spectrum of these states and estimate their mixing strength both in $u,d$ and $s$ quark case to help guide searches for
these states in 1$^{-+}$ sector.

Our methodology is introduced in Section \ref{sec:II}. Then we discuss $1^{++}$ states, $1^{--}$ states, and $1^{-+}$ states in Sections \ref{sec:III}, \ref{sec:IV}, \ref{sec:V} respectively. We discuss the importance of non-perturbative terms in calculations of evaluating mixing strength in Section \ref{sec:VI}. Finally we give our summary and conclusions in the last section.

\section{\label{sec:II}QCDSR approach and mixing strength}
In QCDSR, we normally construct a mixing current combining two state interpretations. The two-point correlation  function of the mixing currents can be written as
\begin{widetext}
\begin{equation} \label{2.1}
\begin{aligned}
\Pi\left ( q^{2} \right )&=i\int d^{4}x\, e^{iqx}\left \langle 0 \left |T\left \{(j_{a}\left ( x \right )+c j_b(x) )(j_{a}^{+}( 0)+ c j_b^{+}(0) )\right \}\right| 0 \right \rangle\\
&=\Pi_a(q^2)+2c\Pi_{ab}(q^2)+c^2\Pi_b(q^2),
\end{aligned}
\end{equation}
\end{widetext}
where $j_a$ and $j_b$ have the same quantum numbers, $c$ is a real parameter related to the mixing strength (not the mixing strength itself, since $c$  may not been normalized), and
\begin{equation}\label{2.2}
\begin{aligned}
\Pi_a\left ( q^{2} \right )&=i\int d^{4}x\, e^{iqx}\left \langle 0 \left |T(j_{a}(x)j_a^{+}(0))\right| 0 \right \rangle,\\
\Pi_b\left ( q^{2} \right )&=i\int d^{4}x\, e^{iqx}\left \langle 0 \left |T(j_{b}(x)j_b^{+}(0))\right| 0 \right \rangle,\\
\Pi_{ab}\left ( q^{2} \right )&=\frac{i}{2}\int d^{4}x\, e^{iqx}\left \langle 0 \left |T(j_{a}(x)j_b^{+}(0)+j_{b}(x)j_a^{+}(0))\right| 0 \right \rangle.
\end{aligned}
\end{equation}

Here we consider the mixed correlator $\Pi_{ab}$ because it provides a signal of which states couple to both currents . One can insert a complete set of particle eigenstates between $j_a$ and $j_b$, and the state with relatively strong coupling to both these currents will be selected out through the QCDSR. By estimating the mass, coupling constants and taking into account experimental results, one can get insight into the constituent composition of the corresponding states. This method worked well in our previous study on vector and scalar meson states~\cite{Chen:2019buw}, and has been successfully applied in other systems~\cite{Harnett:2008cw,Palameta:2017ols,Palameta:2018yce}.

$\Pi_{ab}$ usually can be decomposed into different Lorentz structure
\begin{equation}\label{2.3}
\begin{aligned}
\Pi_{ab}(q^2)&=\sum_n \Pi_{n}(q^2)A_{n}\textrm{,}
\end{aligned}
\end{equation}
where $n=1, 2, 3 ... $, $\Pi_n(q^2)$ is the mixing state correlation function with  specific quantum numbers, and $A_n$ is corresponding Lorentz structure. The forms of $A_n$ are related to $j_a$ and $j_b$, and we will define them in  sections below. For simplicity, we assume a specific $\Pi_{H}(q^2)$ represents a  mixed-state correlation function, is one of the possible  $\Pi_{n}(q^2)$, and that $\Pi_H(q^2)$ obeys dispersion relation~\cite{Shifman:1978bx}
\begin{equation}\label{2.4}
\begin{aligned}
\Pi_{H}(q^2)&=\int_{s_{\textrm{min}}}^{\infty }ds\frac{\rho_{H}(s)}{s-q^{2}-i\epsilon}+\ldots\textrm{,}
\end{aligned}
\end{equation}
where the spectral density $\rho_{H}(s)=\frac{1}{\pi}\textrm{Im}\Pi_{H}(s)$, $s_{\textrm{min}}$ represents the physical threshold  of the corresponding current and dots on the right hand side represent polynomial subtraction terms to render $\Pi_H(q^2)$ finite. The spectral density $\rho_{H}(s)$ can be calculated using the operator product expansion (OPE). In this paper, we calculate the spectral density $\rho_H(s)$ up to  dimension-six operators,
\begin{equation}\label{2.5}
\begin{aligned}
\rho_H^{\textrm{(OPE)}}(s)&=\rho_H^{\textrm{(pert)}}(s)+\rho_H^{\left \langle \bar{q}q \right \rangle}(s)+\rho_H^{\left \langle G^2\right \rangle} (s)\\
&+\rho_H^{\left \langle \bar{q}Gq\right \rangle}(s)+\rho_H^{\left \langle \bar{q}q \right \rangle ^2}(s)+\ldots~\textrm{,}
\end{aligned}
\end{equation}
then
\begin{equation}\label{2.5b}
\Pi_{H}^{\textrm{(OPE)}}(q^2)=\int_{s_{\textrm{min}}}^{\infty }ds\frac{\rho^{\textrm{(OPE)}}_{H}( s)}{s-q^{2}-i\epsilon}+\ldots~.
\end{equation}
On the phenomenological side, by using the narrow resonance spectral density model,
\begin{equation}\label{2.6}
\Pi^{\textrm{(phen)}}_H(q^2)=\frac{\lambda_{a}\lambda_{b}^{*}+\lambda_{b}\lambda_{a}^{*}}{2}\frac{1}{M_H^2-q^2}+\int_{s_{0}}^{\infty }ds\frac{\rho^{\textrm{(cont)}}_{H}(s)}{s-q^{2}-i\epsilon},
\end{equation}
where $\lambda_{a}$ and $\lambda_{b}$ are the  respective couplings of the ground state to the corresponding currents, $M_H$ represents mass of mixed state which has relatively strong coupling to 
the corresponding currents, $\rho_H^{\textrm{(cont)}}$ represents continuum contributions to spectral density,
and $s_0$ is the continuum threshold. By using the QCDSR continuum spectral density assumptions
\begin{equation}\label{2.7}
\begin{aligned}
&\rho_H^{(\textrm{cont})}(s)=\rho_H^{\textrm{(OPE)}}(s)\Theta(s-s_0),
\end{aligned}
\end{equation}
and equating the OPE side and the phenomenological side of the correlation function, $\Pi^{\textrm{(phen)}}_{H}(q^2)=\Pi_{H}^{\textrm{(OPE)}}(q^2)$, we obtain the QCDSR master equation
\begin{equation}\label{2.8}
\int_{s_{\textrm{min}}}^{s_0 }ds\frac{\rho^{\textrm{(OPE)}}_{H}( s)}{s-q^{2}-i\epsilon}+\ldots=\frac{\lambda_{a}\lambda_{b}^{*}+\lambda_{b}\lambda_{a}^{*}}{2}\frac{1}{M_H^2-q^2}.
\end{equation}
After applying the Borel transformation operator $\hat B$ to both side of the master equation,  the subtraction terms are eliminated and the master equation can be written as~\cite{Shifman:1978bx,Shifman:1978by}
\begin{equation} \label{2.9}
\int_{s_{\textrm{min}}}^{s_{0} }ds \,\rho^{\textrm{(OPE)}}_H \left ( s \right )e^{-s\tau}=\frac{\lambda_{a} \lambda_{b}^{*}+\lambda_{a}^* \lambda_{b}}{2}e^{-M^{2}_{H}\tau}\textrm{,}
\end{equation}
where the  Borel parameter $\tau=1/M^2$, and $M$ is the Borel mass. The master equation
(\ref{2.9}) is the foundation of our analysis.
 By taking the $\tau$ logarithmic derivative of Eq.~(\ref{2.9}), we obtain
 \begin{equation}\label{2.10}
M_H^2=\frac{\int_{s_{\textrm{min}}}^{\infty}ds \ s \rho_H^{\textrm{(OPE)}} \left ( s \right )e^{-s\tau}}{\int_{s_{\textrm{min}}}^{\infty}ds \ \rho_H^{\textrm{(OPE)}} \left ( s \right )e^{-s\tau}}.
 \end{equation}
 One can set $a=b$ in Eqs.~(\ref{2.6}), (\ref{2.8}), (\ref{2.9}), and get the original pure state QCDSR.

Because of the OPE truncation and the simplified assumption for the
phenomenological spectral density, Eq.~(\ref{2.9}) is not valid for all values of $\tau$, thus the determination of the sum rule window, in which the validity of (\ref{2.9}) can be established, is very important.
In the literature, different methods are used in the determination of the $\tau$ sum rule window~\cite{Ho:2018cat,Narison:2009vj}.
In this paper, we follow previous similar studies to restrict resonance and high dimensions condensate contributions (HDC), i.e., the resonance part obeys the relation
\begin{equation}\label{2.11}
\frac{\int_{s_{\textrm{min}}}^{s_0}ds \  \rho_H^{\textrm{(OPE)}}\left ( s \right )e^{-s\tau}}{\int_{s_{\textrm{min}}}^{\infty}ds \ \rho_H^{\textrm{(OPE)}} \left ( s \right )e^{-s\tau}}>40\%,
\end{equation}
while HDC (usually $\left \langle \bar{q}q\right \rangle^2$ in molecular systems) obeys the relation
\begin{equation}\label{2.12}
\frac{|\int_{s_{\textrm{min}}}^{\infty}ds \  \rho_H^{\left \langle \bar{q}q\right \rangle^2}\left ( s \right )e^{-s\tau}|}{|\int_{s_{\textrm{min}}}^{\infty}ds \ \rho_H^{\textrm{(OPE)}} \left ( s \right )e^{-s\tau}|}<10\%.
\end{equation}

Furthermore, the value of $s_0$ is also very important in  QCDSR methods. It is often assumed that the threshold satisfies $\sqrt{s_0}=M_H+\Delta_s$, with $\Delta_s\approx$ 0.5\,GeV especially in molecular state QCDSR calculations~\cite{Mo:2014nua,Matheus:2010qxa}.

The approximation $\sqrt{s_0}=M_H+\Delta_s$ can be understood in QCDSR because the parameter $s_0$ separates the ground state and other excited states' contributions to spectral density. Hence one can set $s_0$  less than  the first excitation threshold in case of involving excited state contributions in spectral density, and $\Delta_s$ represents the approximate mass difference between the ground and first excited states. We  assume that the first excited state is approximately  equal to an excited constituent meson and another ground state constituent meson, then we can establish  $s_0$ by comparing the mass difference between ground constituent meson and the first excited constituent meson of the corresponding state (like charmonium and $D$ meson family in our case). We have listed some experimental data for the charmonium and $D$ meson families in Table \ref{t1} and Table \ref{t2}. One can easily find that the mass difference between the ground state and first excited state are all around $0.5^{+0.1}_{-0.1}$ GeV and the fluctuations are all acceptable in QCDSR approach.

\begin{table*}
\begin{ruledtabular}
\begin{center}
\caption{Charmed meson (c$=\pm1$) states, where $q$ represents $u,d$ quark. The
$\bullet$ symbol indicates particles that have confirmed quantum numbers.}
\label{t1} 
\begin{tabular}{ccccc}

PDG name & Possible structure & Ground state & Possible 1st excited state & $\Delta_s/$MeV \\
\hline
$D$            & $\bar{c}\gamma_5q$  & $ \bullet D(1865)$ & $D(2550)$ & $\sim$685 \\$D_1$            & $\bar{c}\gamma_{\mu}\gamma_5q$  & $\bullet D_1(2420)$ & - & -\\
$D_0^*$                                 &  $\bar{c}q$ & $\bullet D_0^*(2300)$ &$D_J^*(2600)$  & $\sim$300 \\
$D^*$          &    $\bar{c}\gamma_{\mu}q$ & $\bullet D^*(2007)$ & $ D^*(2640)$ & $\sim$633\\
\end{tabular}
\end{center}
\end{ruledtabular}
\end{table*}
\begin{table*}
\begin{ruledtabular}
\begin{center}
\caption{Charmonium (possibly non-$\bar{c}c$ states). The $\bullet$ symbol  indicates particles that have confirmed quantum numbers.}
\label{t2} 
\begin{tabular}{cccc}
Possible structure & Ground state/MeV & Possible 1st excited state/MeV & $\Delta_s/$MeV \\
\hline
 $\bar{c}\gamma_5c$  & $ \bullet \eta_c(1S)$/2984 & $ \bullet \eta_c(2S)$/3637 & $\sim$653 \\ $\bar{c}\gamma_{\mu}\gamma_5c$  & $\bullet \chi_{c1}(1P)$/3510 & $\bullet \chi_{c1}(3872)$ & $\sim$362\\
 $\bar{c}c$ & $\bullet \chi_{c0}(1P)$/3415 &$ \chi_{c0}(3860)$ & $\sim$445 \\    $\bar{c}\gamma_{\mu}c$ & $\bullet J/\psi (1S)$/3097 & $\bullet \psi(2S)$/3686 & $\sim$589\\

\end{tabular}
\end{center}
\end{ruledtabular}
\end{table*}

In order to estimate the mixing strength of the physical state strongly coupled to both the two different currents, we define
\begin{equation} \label{2.13}
\begin{aligned}
N\equiv&\frac{|\lambda_{a} \lambda_{b}^{*}+\lambda_{a}^* \lambda_{b}|}{2|\lambda^{'}_{a}\lambda^{'}_{b}|}\textrm{,}
\end{aligned}
\end{equation}
where $\lambda^{'}_{a}$ and $\lambda^{'}_{b}$ are coupling constants of the relevant current with a pure state (i.e., the coupling that emerges in the diagonal correlation functions $\Pi_a$, $\Pi_b$).
Eq.~\eqref{2.13} is analogous to the mixing parameter defined in Ref.~\cite{Hart:2006ps}.  By using appropriate factors of mass  in the definitions of $\lambda^{'}_{a}$ and $\lambda^{'}_{b}$, we
can therefore compare the magnitude of coupling constants and estimate the mixing strength  self-consistently.  The mixing strength depends on the definition and normalization of mixed state. For example, in Ref.~\cite{Narison:1984bv} the definition of the mixed state is
\begin{equation}\label{2.14}
\left |\textrm{M}  \right \rangle=\cos \theta \left |\textrm{A}  \right \rangle+\sin \theta \left |\textrm{B}  \right \rangle ,
\end{equation}
where $\left |\textrm{M}  \right \rangle$ is a mixed state composed of pure states  $\left |\textrm{A} \right \rangle$  and
$\left |\textrm{B} \right \rangle$
and $\theta$ is a mixing angle. In this definition and normalization of the mixed state,
we see that $N\approx \cos \theta \sin \theta$, and $N\in \left ( 0, \frac{1}{2} \right )$.
Because of the different possible normalizations and mixed state definitions, we use  Eq.~\eqref{2.13} as a robust parameter to quantify mixing effects. Furthermore, because the behavior of $N$ is not linear, we define $\widetilde{N}$ under scenario of Eq.(\ref{2.14}),
\begin{equation}\label{2.15}
\widetilde{N}=\sin^2\left(\frac{\arcsin(2N)}{2}\right).
\end{equation}
The quantity $\widetilde{N}$ gives the approximate proportion of pure part of the mixed state. By comparing the mixed state mass with two relevant pure states, suggests that the mixed state is dominated by the part whose pure state mass prediction is closest to the mixed state mass. Otherwise, different decay widths can also help us to distinguish the dominant part of the mixed state.

We use the following numerical values of vacuum condensates consistent with other QCDSR analyses of XYZ states:  $\left \langle \bar{q}q \right \rangle =$ $(-0.23\pm 0.03)^3$GeV$^3$, $\left \langle \bar{q}g_s \sigma G q \right \rangle =$ $m_0^2 \left \langle \bar{q}q \right \rangle$, $m_0^2=$0.8 GeV$^2$,  $\left \langle \alpha_s G^2 \right \rangle =$0.07$\pm0.02$ GeV$^4$, $\left \langle \bar{s}s \right \rangle =$ $(0.8\pm 0.2)\left \langle \bar{q}q \right \rangle $~\cite{Reinders:1984sr,Narison:2010wb}.
In addition, the quark masses $m_c=$1.27 GeV, $m_q$=$\frac{1}{2}(m_u+m_d)$=0.004 GeV, $m_s$= 0.096 GeV, at the energy scale $\mu=$ 2 GeV~\cite{Zyla:2020zbs}, are used.

\section{\label{sec:III}Mixed state in 1$^{++}$ channel}
We start from the following three forms of currents:
\begin{equation}\label{++}
\begin{aligned}
j^{X_A}_{\mu\nu}(x)=&\frac{i}{\sqrt{2}}[\bar{c}(x)\gamma_{\mu}c(x) \bar{q}(x)\gamma_{\nu}q(x)\\
-&\bar{c}(x)\gamma_{\nu}c(x) \bar{q}(x)\gamma_{\mu}q(x)], \\
j^{X_{B_1}}_{\mu}(x)=&\frac{i}{\sqrt{2}}[\bar{c}(x)\gamma_{\mu}q(x)\bar{q}(x)\gamma_{5}c(x)\\
-&\bar{q}(x)\gamma_{\mu}c(x)\bar{c}(x)\gamma_{5}q(x)],\\
j^{X_{B_2}}_{\mu}(x)=&\frac{i}{\sqrt{2}}[\bar{c}(x)\gamma_{\mu}\gamma_5q(x)\bar{q}(x)c(x)\\
+&\bar{q}(x)\gamma_{\mu}\gamma_5c(x)\bar{c}(x)q(x)],\\
j^{X_{C}}_{\mu}(x)=&\frac{1}{6\sqrt{2}}\left \langle \bar{q}q \right \rangle\bar{c}(x)\gamma_{\mu}\gamma_5c(x),
\end{aligned}
\end{equation}
 where $X$ denotes the 1$^{++}$ state,  the subscript $A$ of $X$ denotes the $\bar{Q}Q\bar{q}q$ scenario, $B$ denotes the $\bar{Q}qQ\bar{q}$ scenario, $C$ denotes the $\bar{Q}Q$ scenario, and the corresponding mesonic structures  of these currents are listed in Table \ref{t3}. We note that the former two currents can be decomposed into two constituent meson currents, and the mass prediction of each corresponding pure state are usually close to the sum of masses of these two constituent mesons. The current $j^{X_A}_{\mu\nu}$ can be decomposed into $J/\psi(1S)(3097)$ and $\rho(770)$ currents, $j^{X_{B_1}}_{\mu}(x)$ can be decomposed into  $D^*(2007)$ and $D(1865)$ currents. The sum of masses of two constituent mesons are both close to $X(3872)$. Hence we choose these two currents to study $X(3872)$. The current $j^{X_{B_2}}_{\mu}$ has the same quantum numbers and it cannot be excluded in $1^{++}$ mixing state structures. Besides, $j^{X_{C}}_{\mu}(x)$ is normalized according to Ref.\, \cite{Matheus:2009vq}. Since the mixing between $j^{X_{C}}_{\mu}(x)$ and $\bar{Q}Q\bar{q}q$  is suppressed ( $\bar{q}q$ in $\bar{Q}Q\bar{q}q$ becomes a bubble and vanishes), we only consider the mixing between $j^{X_{C}}_{\mu}(x)$ and $j^{X_{B_1}}_{\mu}(x)$ or $j^{X_{B_2}}_{\mu}(x)$. 

To study the pure $\bar{Q}Q\bar{q}q$, $\bar{Q}qQ\bar{q}$ and $\bar{Q}Q $ states, the respective two-point correlation functions  can be decomposed into different Lorentz structures
\begin{equation}
\begin{aligned}
\Pi_{\mu\nu\rho\sigma}^{X_A}(q^{2})=&\Pi_{a}^{X_A}(q^2)\frac{1}{q^2}(q^2g_{\mu\rho}g_{\nu\sigma}-q^2g_{\mu\sigma}g_{\nu\rho}\\
-&q_{\mu}q_{\rho}g_{\nu\sigma}+q_{\mu}q_{\sigma}g_{\nu\rho}-q_{\nu}q_{\sigma}g_{\mu\rho}+q_{\nu}q_{\rho}g_{\mu\sigma})\\
+&\Pi_{b}^{X_A}(q^2)\frac{1}{q^2}(-q_{\mu}q_{\rho}g_{\nu\sigma}+q_{\mu}q_{\sigma}g_{\nu\rho}\\
-&q_{\nu}q_{\sigma}g_{\mu\rho}+q_{\nu}q_{\rho}g_{\mu\sigma}),\\
\Pi_{\mu\nu}^{X_{B_k}/C}(q^{2})=&\Pi_{(1)}^{X_{B_k}/C}(q^2) \left (-g_{\mu\nu}+\frac{q_{\mu}q_{\nu}}{q^2} \right )\\
+&\Pi_{(0)}^{X_{B_k}/C}(q^2) \left (\frac{q_{\mu}q_{\nu}}{q^2} \right ),
\end{aligned}
\end{equation}
where $k=1,2$, $\Pi_{a}^{X_A}$ and $\Pi_{b}^{X_A}$ describes pure molecular state contribution with respective quantum numbers $1^{++}$ and $1^{-+}$, $\Pi_{(1)}^{X_{B_k}/C}$ and $\Pi_{(0)}^{X_{B_k}/C}$ respectively describes $1^{++}$ and $0^{-+}$ state contributions.
In the mixing scenario, we start from the off-diagonal mixed
correlator described in the previous section, i.e.,
\begin{equation}\label{++2}
\begin{aligned}
\Pi^{M_{X_1}}_{\mu\nu\sigma}(q^{2})=&\frac{i}{2}\int d^{4}x \,e^{iq\cdot x}  \langle 0|
T(j^{X_A}_{\mu\nu}(x)j^{X_{B_1}+}_{\sigma}(0)\\
+&j^{X_{B_1}}_{\sigma}(x)j^{X_{A}+}_{\mu\nu}(0) ) |0 \rangle, \\
\Pi^{M_{X_2}}_{\mu\nu\sigma}(q^{2})=&\frac{i}{2}\int d^{4}x \,e^{iq\cdot x}  \langle 0|
T(j^{X_A}_{\mu\nu}(x)j^{X_{B_2}+}_{\sigma}(0)\\
+&j^{X_{B_2}}_{\sigma}(x)j^{X_{A}+}_{\mu\nu}(0) ) |0 \rangle,
\end{aligned}
\end{equation}
where $M_{X_k}$, $k=$1,2, are mixed states assumed to result from the corresponding currents. These
mixed correlators have the Lorentz structure
\begin{equation}
\begin{aligned}
&\Pi_{\mu\nu\sigma}^{M_{X_k}}(q^{2})=\Pi^{M_{X_k}}(q^{2})(q_{\alpha}\epsilon_{\alpha\mu\nu\sigma}).\\
\end{aligned}
\end{equation}
Besides, when we consider the two-quark states $j^{X_{C}}_{\mu}(x)$, the mixed correlator and its Lorentz structure are
\begin{equation}
\begin{aligned}
\Pi^{M_{C}}_{\mu\nu}(q^{2})=&\frac{i}{2}\int d^{4}x \,e^{iq\cdot x}  \langle 0|
T(j^{X_C}_{\mu}(x)j^{X_{B_1}+}_{\nu}(0)\\
+&j^{X_{B_1}}_{\nu}(x)j^{X_C+}_{\mu}(0) ) |0 \rangle, \\
\Pi_{\mu\nu}^{M_{C}}(q^{2})=&\Pi_{(1)}^{M_{C}}(q^2)\left(-g_{\mu\nu}+\frac{q_{\mu}q_{\nu}}{q^2}\right)\\
+&\Pi_{(0)}^{M_{C}}(q^2)\left(\frac{q_{\mu}q_{\nu}}{q^2}\right),
\end{aligned}
\end{equation}
where $\Pi_{(1)}^{C}$ and $\Pi_{(0)}^{C}$ respectively describes $1^{++}$ and $0^{-+}$ state contributions. Here we just consider the state mixed with $j^{X_{C}}_{\mu}$ and $j^{X_{B_1}}$, which is a candidate to $X(3872)$.

We follow the $40\%-10\%$ sum-rule window and $\Delta_s$ methods mentioned in Section \ref{sec:II}. After establishing $\tau$ window by the $40\%-10\%$ method at a specific $s_0$, for each state we 
use Eq.~(\ref{2.10}) to plot the $\tau$ behavior of $M_H$ for the chosen $s_0$ 
(see Appendix \ref{Appendix:A} for details). The mass prediction 
$M_H$ and $s_0$ are then compared with the constraint $\sqrt{s_0}=M_H+\Delta_s$, then $s_0$ is adjusted and the analysis is repeated until we find the
best $(M_H,s_0)$ solutions satisfying the relation $\sqrt{s_0}=M_H+\Delta_s$. The coupling constants are naturally obtained through the predicted $M_H$ and $s_0$ according to Eq.~(\ref{2.9}). The mass and coupling constant prediction and associated QCDSR parameters are collected in Table \ref{t3}. All the parameters are the average values in the corresponding $\tau$ window.

The uncertainties mainly  are from the input parameters .  For instance $\left \langle \alpha_sG^2 \right \rangle =$ $0.07\pm 0.02 $GeV$^4$,$\left \langle \bar{s}s \right \rangle =$ $(0.8\pm 0.2)\left \langle \bar{q}q \right \rangle $ and  $\left \langle \bar{q}q \right \rangle =$ $(-0.23\pm 0.03)^3$GeV$^3$. The  quark masses and other parameters we have included in calculations have less than 5\% uncertainties due to substantial  numerical fittings by other researchers. There is also  an uncertainty about the value of the threshold $s_0$. Analogue to the studies in Refs.\cite{Wang:2014gwa,Zhang:2009em}, the fluctuation of threshold is set to be $\pm 0.1$GeV($\sqrt s_0$).

In the pure state calculations, a $\tau$ window of $X_A$(3798) state cannot be determined under $40\%-10\%$ and $\Delta_s$ method and we rearrange the limits of resonance and HDC to $35\%-15\%$ (one may naturally expect that the pure-state analysis requires such adjustments because of mixing ). The states $X_A$(3798), $X_{B_1}$(3857) both have mass predictions close to $X(3872)$. However, the large mass prediction of the $X_{B_2}$(5310)  state is much beyond the $D_1$+$D_0^*$ threshold,and does not match the observed $1^{++}$ states.
\begin{table*}
\begin{ruledtabular}
\begin{center}
\caption{Summary of results for $1^{++}$ states. $\lambda=\frac{\lambda_a\lambda_b^*+\lambda_a^*\lambda_b}{2}$ when mixed cases involved, the same below.}
\label{t3} 
\begin{tabular}{cccccc}

State & Current structure & Mass/GeV & $\lambda$/$10^{-4}$ GeV$^{10}$ & $\sqrt{s_0}$/GeV & $\tau$ window/GeV$^{-2}$  \\
\hline
$X_A$& $J/\psi \rho$  & $3.798^{+0.09}_{-0.09}$ &  $1.49^{+0.51}_{-0.47}$  & 4.4 &0.30 -- 0.31 \\
$X_{B_1}$            & $D^*\bar{D}$  & $3.857^{+0.06}_{-0.06}$ & $2.24^{+0.65}_{-0.53}$ & 4.4 & 0.31 -- 0.39\\
$X_{B_2}$            & $D_1\bar{D}_0^*$  & $5.310^{+0.04}_{-0.04}$ & $69.0^{+15.0}_{-14.0}$ & 5.8 & 0.20 -- 0.29 \\
$X_C$            &  $\chi_{c1}$ & $3.511^{+0.02}_{-0.03}$ & $0.0229^{+0.0017}_{-0.0018}$ & 4.5 & 0.29 -- 0.31 \\
$M_{X_{1}}$   &  $J/\psi \rho$ - $D^*\bar{D}$ & $3.987^{+0.06}_{-0.06}$ & $0.168^{+0.049}_{-0.042}$ GeV$^{-1}$ & 4.4 & 0.30 -- 0.32 \\
$M_{X_{2}}$   &  $J/\psi \rho$ -- $D_1\bar{D}_0^*$ & $4.945^{+0.08}_{-0.06}$  & $0.760^{+0.29}_{-0.20}$ GeV$^{-1}$ & 5.45 & 0.22 -- 0.24\\
$M_{C}$   &  $\chi_{c1}$ - $D^*\bar{D}$ & $3.818^{+0.03}_{-0.02}$  & $0.0282^{+0.0027}_{-0.0024}$ & 4.5 & 0.28 -- 0.30\\

\end{tabular}
\end{center}
\end{ruledtabular}
\end{table*}

The mixing strength can then be estimated by computing the value of $N$ via Eq.~\eqref{2.13}. Note that the coupling constants of two mixed state correlators have the form
\begin{equation}
\begin{aligned}\label{struc22}
\Pi^{M_{X_k}}_{\mu\nu\sigma}(q^{2})&=\frac{i}{2}\int d^{4}x \,e^{iq\cdot x}  \langle 0|
T(j^{X_A}_{\mu\nu}(x)j^{X_{B_k}+}_{\sigma}(0)\\
&+j^{X_{B_k}}_{\sigma}(x)j^{X_{A}+}_{\mu\nu}(0) ) |0 \rangle\\ &\sim (\frac{\lambda_{X_A}}{M_H}\epsilon_{\mu\nu\alpha\beta}\varepsilon^{X_A}_{\alpha}q_{\beta})(\lambda^*_{X_{B_k}}\varepsilon^{*X_{B_k}}_{\sigma})\\
&+(\frac{\lambda_{X_A}^*}{M_H}\epsilon_{\mu\nu\alpha\beta}\varepsilon^{*X_A}_{\alpha}q_{\beta})(\lambda_{X_{B_k}}\varepsilon^{X_{B_k}}_{\sigma})+...\\
&=\frac{\lambda_{X_A}\lambda^*_{X_{B_k}}+\lambda^*_{X_A}\lambda_{X_{B_k}}}{M_H}q_{\alpha}\epsilon_{\mu\nu\sigma\alpha }+...,
\end{aligned}
\end{equation}
where $k=1,2$, $\varepsilon_{\alpha/\sigma}^{X_A/X_{B_k}}$ is polarization vector, $M_H$ represents the ground state mass of $X_A$, dots represent excited contributions to the spectral density and polynomial subtraction terms. In the definition of the mixing strength Eq.~(\ref{2.13}), we have omitted the Lorentz structures of corresponding currents. 
The dimension of the decay constant depends on  the Lorentz structure  we extract in the diagonal correlator . If 
the two currents have different Lorentz structure, we need to compensate the mass dimension of the decay constants 
, which are obtained from other works, to make the mixing strength Eq.~(\ref{2.13}) dimensionless. The normal method 
is to make the Lorentz structures  massless by multipling a factor $ M^n_H$ with a suitable $n$. 
 Hence we define $\lambda_{X_A}/M_H$ as the new coupling constant of $X_A$ state .  The mixing strength can be written as
\begin{equation}\label{firstn}
\begin{aligned}
&N_{M_{X_1}}=\frac{0.168\textrm{GeV}^9\times M_{H}(3.798 \textrm{GeV})}{\sqrt{1.49}\textrm{GeV}^5\times\sqrt{2.24}\textrm{GeV}^5}=0.349,\\
&\widetilde{N}_{M_{X_1}}=\sin^2\left(\frac{\arcsin(0.349\times2)}{2}\right)=14\%,\\
&N_{M_{X_2}}=\frac{0.760\textrm{GeV}^9\times M_{H}(3.798 \textrm{GeV})}{\sqrt{1.49}\textrm{GeV}^5\times\sqrt{69.0}\textrm{GeV}^5}=0.285,\\
&\widetilde{N}_{M_{X_2}}=\sin^2\left(\frac{\arcsin(0.285\times2)}{2}\right)=9.0\%,\\
&N_{M_C}=\frac{0.0282\textrm{GeV}^{10}}{\sqrt{0.0229}\textrm{GeV}^5\times\sqrt{2.24}\textrm{GeV}^5}=0.125,\\
&\widetilde{N}_{M_{C}}=\sin^2\left(\frac{\arcsin(0.125\times2)}{2}\right)=1.6\%,\\
\end{aligned}
\end{equation}

The state $M_{X_1}$(3987) is a mixture of $X_A$(3798) and $X_{B_1}$(3857) which have similar mass predictions close to $X(3872)$, and unsurprisingly has the same mass prediction. Due to $X(3872)$ observed  decays to $\pi^+\pi^-J/\psi(1S)$, $\omega J/\psi(1S)$ and $\bar{D}^{*0}D^0$, $M_{X_1}$(3987) is a good candidate to describe the $X(3872)$~\cite{Zyla:2020zbs}. We can estimate proportions of each constituent and decay width of corresponding decay modes by using the parameter $N_{M_{X_1}}$. Experimental results of $X(3872)$ decay width $\Gamma_1$ of $\bar{Q}q+Q\bar{q}$ like decay mode is $>30\%$, while decay width $\Gamma_2$ of $\bar{Q}Q+\bar{q}q$ like decay mode is $>5\%$.By comparison, the parameter $N_{M_{X_1}}$ shows that the
proportions of 
the $\bar{Q}qQ\bar{q}$ and $\bar{Q}Q\bar{q}q$ parts of $M_{X_1}$(3987) are respectively 86$\%$ and 14$\%$. Considering similar Lorentz-invariant phase-space of these two kinds of decay modes, we can roughly equate $\Gamma_1/\Gamma_2$ to the ratio of these two parts, $86\%/14\%\sim 6$, which is consistent with experimental results. 
It should be noted that our method  can not determine  definitely which constitute dominates the mixing state. We tend to 
the one whose pure mass is closer to the mixing state.

When we consider the two-quark state $j^{X_{C}}_{\mu}$, the corresponding mixing angle is $\text{arcsin(0.25)}/2=7^{\circ}$, which is consist with the result in Ref.\,\cite{Matheus:2009vq}, and the dominant part of $M_C$ is $j^{X_{C}}_{\mu}$. However, we found that this result strongly depends on the normalization of $j^{X_{C}}_{\mu}$. Hence a proper normalized current is essential in calculations. 

For the state $M_{X_2}$(4945), it's mass prediction is larger than all observed $1^{++}$ states. However, our calculation suggests that it is relatively strongly mixed . The dominant part of $M_{X_2}$(4945) is more likely to be $X_{B_2}$(5310) by comparing mass predictions.

In Ref.~\cite{Wang:2013daa}, authors have calculated state $X_{B_1}$ with similar method, and got results: the mass $m_{X_{B_1}}=3.89^{+0.09}_{-0.09}$ GeV, the decay constant $\lambda_{X_{B_1}}=2.96^{+1.09}_{-0.79}\times 10^{-4}$ GeV$^{10}$ with $\sqrt{s_0}=4.41$ GeV, which is consistent with our results.

\section{\label{sec:IV}Mixed state in $1^{--}$ channel}
We start from two forms of  currents as follows:
\begin{equation}
\begin{aligned}
j^{Y_{A_1}/Y_{A_{s1}}}_{\mu}(x)=&\bar{c}(x)c(x) \bar{q}(x)\gamma_{\mu}q(x), \\
j^{Y_{A_2}/Y_{A_{s2}}}_{\mu}(x)=&\bar{c}(x)\gamma_{\mu}c(x) \bar{q}(x)q(x), \\
j^{Y_{B_1}/Y_{B_{s1}}}_{\mu}(x)=&\frac{i}{\sqrt{2}}[\bar{c}(x)\gamma_{\mu}q(x)\bar{q}(x)c(x)\\
+&\bar{q}(x)\gamma_{\mu}c(x)\bar{c}(x)q(x)],\\
j^{Y_{B_2}/Y_{B_{s2}}}_{\mu}(x)=&\frac{i}{\sqrt{2}}[\bar{c}(x)\gamma_{\mu}\gamma_5q(x)\bar{q}(x)\gamma_5c(x)\\
-&\bar{q}(x)\gamma_{\mu}\gamma_5c(x)\bar{c}(x)\gamma_5q(x)],
\end{aligned}
\end{equation}
where $Y$ denotes the $1^{--}$ state, the subscript $A$ of $Y$ represents $\bar{Q}Q\bar{q}q$ scenario while $B$ represents the $\bar{Q}qQ\bar{q}$ scenario, the additional subscript $s$ represents the $s$ quark case, and one can straightforwardly replace $q$ with $s$ quark when $j^{Y_{A_{s1}}}_{\mu}$, $j^{Y_{A_{s2}}}_{\mu}$, $j^{Y_{B_{s1}}}_{\mu}$ and $j^{Y_{B_{s2}}}_{\mu}$ are involved. The $Y(4230)$ was observed to decay to $\chi_{c0}\omega$, while $Y(4660)$ was observed to have both $\psi(2S)\pi^+\pi^-$ and $D_s^+D_{s1}(2536)^-$ decay modes~\cite{Zyla:2020zbs}. Hence we especially focus on   currents $j^{Y_{A_1}}_{\mu}$ and  $j^{Y_{A_{s2}}}_{\mu}$ which are consistent with the respective decay modes, to describe $Y(4230)$ and $Y(4660)$ respectively, and discuss the corresponding mixed states in both $u,d$ quark and $s$ quark for simplicity.

The two-point correlator functions of pure states have Lorentz structures
\begin{equation}
\begin{aligned}
\Pi_{\mu\nu}^{Y_{A_k}/Y_{A_{sk}}}(q^{2})=&\Pi_{(1)}^{Y_{A_k}/Y_{A_{sk}}}(q^2)\left(-g_{\mu\nu}+\frac{q_{\mu}q_{\nu}}{q^2}\right)\\
+&\Pi_{(0)}^{Y_{A_k}/Y_{A_{sk}}}(q^2)\left(\frac{q_{\mu}q_{\nu}}{q^2}\right),\\
\Pi_{\mu\nu}^{Y_{B_{k}}/Y_{B_{sk}}}(q^{2})=&\Pi_{(1)}^{Y_{B_k}/Y_{B_{sk}}}(q^2)\left(-g_{\mu\nu}+\frac{q_{\mu}q_{\nu}}{q^2}\right)\\
+&\Pi_{(0)}^{Y_{B_k}/Y_{B_{sk}}}(q^2)\left(\frac{q_{\mu}q_{\nu}}{q^2}\right),
\end{aligned}
\end{equation}
where $k=$1,2, $\Pi_{(1)}^{Y_{A_k}/Y_{A_{sk}}}$ and $\Pi_{(1)}^{Y_{B_k}/Y_{B_{sk}}}$ describe pure state contributions with quantum numbers $1^{--}$,  $\Pi_{(0)}^{Y_{A_k}/Y_{A_{sk}}}$ and $\Pi_{(0)}^{Y_{B_k}/Y_{B_{sk}}}$ describe pure state contribution with quantum numbers $0^{+-}$.

To study the mixed state, the off-diagonal mixing two-point correlation functions described in Section \ref{sec:II} are
\begin{equation}
\begin{aligned}
\Pi^{M_{Y_1}}_{\mu\nu}(q^{2})=&\frac{i}{2}\int d^{4}x \,e^{iq\cdot x}  \langle 0|
T(j^{Y_{A_1}}_{\mu}(x)j^{Y_{B_1}+}_{\nu}(0)\\
+&j^{Y_{B_1}}_{\nu}(x)j^{Y_{A_1}+}_{\mu}(0) ) |0 \rangle, \\
\Pi^{M_{Y_2}}_{\mu\nu}(q^{2})=&\frac{i}{2}\int d^{4}x \,e^{iq\cdot x}  \langle 0|
T(j^{Y_{A_1}}_{\mu}(x)j^{Y_{B_2}+}_{\nu}(0)\\
+&j^{Y_{B_2}}_{\nu}(x)j^{Y_{A_1}+}_{\mu}(0) ) |0 \rangle, \\
\Pi^{M_{Y_{s1}}}_{\mu\nu}(q^{2})=&\frac{i}{2}\int d^{4}x \,e^{iq\cdot x}  \langle 0|
T(j^{Y_{A_{s2}}}_{\mu}(x)j^{Y_{B_{s1}}+}_{\nu}(0)\\
+&j^{Y_{B_{s1}}}_{\nu}(x)j^{Y_{A_{s2}}+}_{\mu}(0) ) |0 \rangle, \\
\Pi^{M_{Y_{s2}}}_{\mu\nu}(q^{2})=&\frac{i}{2}\int d^{4}x \,e^{iq\cdot x}  \langle 0|
T(j^{Y_{A_{s2}}}_{\mu}(x)j^{Y_{B_{s2}}+}_{\nu}(0)\\
+&j^{Y_{B_{s2}}}_{\nu}(x)j^{Y_{A_{s2}}+}_{\mu}(0) ) |0 \rangle, \\
\end{aligned}
\end{equation}
where $M_{Y_k}$ and $M_{Y_{sk}}$, $k=1,2$, both represent mixed states coupled to their respective currents. These mixed correlators have same Lorentz structures as pure  state cases,
\begin{equation}
\begin{aligned}
\Pi_{\mu\nu}^{M_{Y_{k}}/M_{Y_{sk}}}(q^{2})=&\Pi_{(1)}^{M_{Y_{k}}/M_{Y_{sk}}}(q^2)\left(-g_{\mu\nu}+\frac{q_{\mu}q_{\nu}}{q^2}\right)\\
+&\Pi_{(0)}^{M_{Y_{k}}/M_{Y_{sk}}}(q^2)\left(\frac{q_{\mu}q_{\nu}}{q^2}\right),
\end{aligned}
\end{equation}
where $k=$1,2, $\Pi_{(1)}^{M_{Y_{k}}/M_{Y_{sk}}}$ and $\Pi_{(0)}^{M_{Y_{k}}/M_{Y_{sk}}}$ respectively describe mixed state with quantum number  $1^{--}$, $0^{+-}$.

We follow same method mentioned in  $1^{++}$ channel. The mesonic structures, mass and coupling constant predictions, and the related QCDSR parameters are collected in Table \ref{t4}.

\begin{table*}
\begin{ruledtabular}
\begin{center}
\caption{Summary of results for $1^{--}$   states.}
\label{t4} 
\begin{tabular}{cccccc}
State & Current structure & Mass/GeV & $\lambda$/$10^{-4}$GeV$^{10}$& $\sqrt{s_0}$/GeV & $\tau$ window/GeV$^{-2}$ \\
\hline
$Y_{A_1}$& $\chi_{c0}\omega$  & $4.207^{+0.08}_{-0.09}$ & $1.64^{+0.63}_{-0.49}$ & 4.8  &  0.27 -- 0.28\\
$Y_{A_{s2}}$            & $J/\psi f(980)$  & $4.621^{+0.05}_{-0.06}$ & $21.3^{+5.2}_{-4.7}$ & 5.1 & 0.25 -- 0.32 \\
$Y_{B_1}$            & $D_0^*\bar{D}^*$  & $4.922^{+0.04}_{-0.04}$ & $34.5^{+6.7}_{-6.0}$ & 5.4 & 0.21 -- 0.34\\
$Y_{B_2}$            & $D_1\bar{D}$  & $4.385^{+0.06}_{-0.06}$ & $7.36^{+1.95}_{-1.67}$ & 4.9 & 0.27 -- 0.35\\
$Y_{B_{s1}}$            & $D_{s0}^*\bar{D}_s^*$  & $4.952^{+0.03}_{-0.04}$ & $37.9^{+6.5}_{-6.3}$ & 5.45 & 0.21 -- 0.36 \\
$Y_{B_{s2}}$            & $D_{s1}\bar{D}_s$  & $4.494^{+0.08}_{-0.05}$ & $9.60^{+3.7}_{-2.1}$ & 5.0 & 0.26 -- 0.39 \\
$M_{Y_{1}}$   &  $\chi_{c0}\omega$ - $D_0^*\bar{D}^*$ & $4.770^{+0.07}_{-0.06}$  & $1.16^{+0.37}_{-0.28}$  & 5.3 & 0.24 -- 0.25\\
$M_{Y_{2}}$   &  $\chi_{c0}\omega$ - $D_1\bar{D}$ & $4.266^{+0.08}_{-0.08}$ & $0.373^{+0.117}_{-0.093}$ & 4.95 & 0.26 -- 0.27\\
$M_{Y_{s1}}$   &  $J/\psi f(980)$ - $D_{s0}^*\bar{D}_s^*$ & $4.610^{+0.05}_{-0.06}$ & $2.56^{+0.67}_{-0.56}$ & 5.1 & 0.24 -- 0.33\\
$M_{Y_{s2}}$   &  $J/\psi f(980)$ - $D_{s1}\bar{D}_s$ & $4.450^{+0.05}_{-0.06}$ & $1.64^{+0.43}_{-0.22}$  & 4.95 & 0.26 -- 0.33 \\
\end{tabular}
\end{center}
\end{ruledtabular}
\end{table*}
In $Y$ family of states, $Y(4160)$, $Y(4260)$, $Y(4415)$, $Y(4660)$ are reported have decay modes including a $s$ quark in final states, and $Y(4230)$, $Y(4360)$, $Y(4390)$ have not been observed to have decay modes including a $s$ quark in final states. Furthermore, $Y(4260)$ only decays to $K$ meson while $Y(4415)$ and $Y(4660)$ only decays to $D_s$ meson when a $s$ quark is directly involved in final states. The $Y(4160)$ has both decay modes including $K$ and $D_s$ mesons in final states. On the other hand, all $Y$ states have both $\bar{Q}Q+\bar{q}q$ like decay modes and $\bar{Q}q+Q\bar{q}$ like decay modes except $Y(4390)$. The decay mode $Y(4390)$  to $\pi^+\pi^-h_c$ were observed, but other decay modes of  $Y(4390)$ have not been seen yet. Because the $K$ meson may decay to $\pi$ meson and disappear in final states, we cannot exclude a $s$ quark in $Y(4230)$, $Y(4360)$, $Y(4390)$~\cite{Zyla:2020zbs}. Hence we suggest that $Y(4230)$ has candidates $Y_{A_1}$(4207), $M_{Y_2}$(4266), $Y(4360)$ or $Y(4390)$ has candidate $Y_{B_2}$(4385), $Y(4415)$ has candidates $Y_{B_{s2}}$(4494) and $M_{Y_{s2}}$(4450), $Y(4660)$ has candidates $Y_{A_{s2}}$(4621) and $M_{Y_{s1}}$(4610). Although the remaining states are not compatible with known $1^{--}$ states, they still possibly mix with other states, and their contributions can be estimated.

For the  $1^{--}$ states, the mixing strengths are given by the data in Table \ref{t4},
\begin{equation}\label{firstn22}
\begin{aligned}
&N_{M_{Y_{1}}}=\frac{1.16\textrm{GeV}^{10}}{\sqrt{1.64}\textrm{GeV}^5\times\sqrt{34.5}\textrm{GeV}^5}=0.15,\\
&N_{M_{Y_{2}}}=\frac{0.373\textrm{GeV}^{10}}{\sqrt{1.64}\textrm{GeV}^5\times\sqrt{7.36}\textrm{GeV}^5}=0.11,\\
&N_{M_{Y_{s1}}}=\frac{2.56\textrm{GeV}^{10}}{\sqrt{21.3}\textrm{GeV}^5\times\sqrt{37.9}\textrm{GeV}^5}=0.09,\\
&N_{M_{Y_{s2}}}=\frac{1.64\textrm{GeV}^{10}}{\sqrt{21.3}\textrm{GeV}^5\times\sqrt{9.60}\textrm{GeV}^5}=0.11.
\end{aligned}
\end{equation}
All mixed states have much weaker mixing strength compared with $1^{++}$ mixed states. We suggest that $1^{--}$ states are preferred to be pure state and weakly mixed with other states. This becomes more clear when we convert $N$ to $\widetilde{N}$,
\begin{equation}\label{ntilda2}
\begin{aligned}
&\widetilde{N}_{M_{Y_1}}=\sin^2\left(\frac{\arcsin(0.15\times2)}{2}\right)=2.3\%,\\
&\widetilde{N}_{M_{Y_2}}=\sin^2\left(\frac{\arcsin(0.11\times2)}{2}\right)=1.2\%,\\
&\widetilde{N}_{M_{Y_{s1}}}=\sin^2\left(\frac{\arcsin(0.09\times2)}{2}\right)=0.82\%,\\
&\widetilde{N}_{M_{Y_{s2}}}=\sin^2\left(\frac{\arcsin(0.11\times2)}{2}\right)=1.2\%,
\end{aligned}
\end{equation}
where the values of $\widetilde{N}$ suggest that the assumed mixed states with quantum numbers $1^{--}$ are actually very pure state. As mentioned above, $M_{Y_1}$(4770) which contains no $s$ quark, is close to $Y(4660)$, cannot be compatible with known $1^{--}$ states. $M_{Y_2}$(4266) which is a possible candidate for $Y(4230)$, is a mixture of $Y_{A_1}$(4207) and $Y_{B_2}$(4385). By comparing the two mass predictions, $M_{Y_2}$(4266) is closer to $Y_{A_1}$(4207) rather than $Y_{B_2}$(4385), and it is possibly dominated by $\bar{Q}Q\bar{q}q$ component. For the same reasons, $M_{Y_{s1}}(4610)$ is possibly dominated by a $\bar{Q}Q\bar{q}q$ component while $M_{Y_{s2}}$(4450) is possibly dominated by $\bar{Q}qQ\bar{q}$. Hence we suggest that $Y(4230)$, $Y(4660)$ prefer a $\bar{Q}Q\bar{q}q$ state, and $Y(4415)$ prefers a $\bar{Q}qQ\bar{q}$ state.

 In Ref.~\cite{Wang:2016wwe}, authors have calculated states $Y_{B_1}$ and $Y_{B_2}$ with similar method and obtained the results:  mass $m_{Y_{B_1}}=4.78^{+0.07}_{-0.07}$ with  $\sqrt{s_0}=5.3$ GeV$^{-2}$, and mass $m_{Y_{B_2}}=4.36^{+0.08}_{-0.08}$ with  $\sqrt{s_0}=4.9$ GeV$^{-2}$, which are consistent with our results. The small difference of mass of $Y_{B_1}$ is caused by different values of $\sqrt{s_0}$. Besides, in Ref.~\cite{Wang:2016wwe}, authors have discussed different results of similar states of $Y_{B_1}$ and $Y_{B_2}$ in previous papers. For instance, authors in Ref.~\cite{Zhang:2009em} did not distinguish the charge conjugations and obtained mass of a $Y_{B_1}$ like state $m_{D^*_0\bar{D^*}}=4.26$ GeV. Our results are more supportive to the results in Ref.~\cite{Wang:2016wwe}.
  
 In Ref.~\cite{Albuquerque:2011ix}, authors have calculated states $Y_{A_{s2}}$ with similar method and obtained the result:  mass $m_{Y_{A_{s2}}}=4.67^{+0.09}_{-0.09}$ with  $\sqrt{s_0}=5.1$ GeV$^{-2}$, which is consistent with our result.
\section{ Mixed state in $1^{-+}$ channel}\label{sec:V}
We start from two forms of  currents as follows:
\begin{equation}
\begin{aligned}
j^{P_A}_{\mu\nu}(x)&=j^{X_A}_{\mu\nu}(x), \\
j^{P_{A_s}}_{\mu\nu}(x)&=\frac{i}{\sqrt{2}}[\bar{c}(x)\gamma_{\mu}c(x) \bar{s}(x)\gamma_{\nu}s(x)\\
&-\bar{c}(x)\gamma_{\nu}c(x) \bar{s}(x)\gamma_{\mu}s(x)], \\
j^{P_{B_1}/P_{B_{s1}}}_{\mu}(x)&=\frac{i}{\sqrt{2}}[\bar{c}(x)\gamma_{\mu}q(x)\bar{q}(x)c(x)\\
&-\bar{q}(x)\gamma_{\mu}c(x)\bar{c}(x)q(x)],\\
j^{P_{B_2}/P_{B_{s2}}}_{\mu}(x)&=\frac{i}{\sqrt{2}}[\bar{c}(x)\gamma_{\mu}\gamma_5q(x)\bar{q}(x)\gamma_5c(x)\\
&+\bar{q}(x)\gamma_{\mu}\gamma_5c(x)\bar{c}(x)\gamma_5q(x)],
\end{aligned}
\end{equation}
where $P$ denotes the $1^{-+}$ state, the subscript $A$ of $P$ represents the $\bar{Q}Q\bar{q}q$ scenario while $B$ represents the $\bar{Q}qQ\bar{q}$ scenario, the additional subscript $s
$ represents the $s$ quark case, and one can straightforwardly replace $q$ with $s$ when  $j^{P_{B_{s1}}}_{\mu}$ and $j^{P_{B_{s2}}}_{\mu}$ are involved. The structures of these currents are similar to the $1^{++}$ and $1^{--}$ cases, and it is interesting to compare the mass predictions of these currents to $1^{++}$ and $1^{--}$ states.

To study the pure $\bar{Q}Q\bar{q}q$ and $\bar{Q}qQ\bar{q}$  states, the two-point correlation functions respectively have Lorentz structure
\begin{equation}
\begin{aligned}
\Pi_{\mu\nu\rho\sigma}^{P_A/P_{A_s}}(q^{2})=&\Pi_{a}^{P_A/P_{A_s}}\frac{1}{q^2}(q^2g_{\mu\rho}g_{\nu\sigma}-q^2g_{\mu\sigma}g_{\nu\rho}\\
-&q_{\mu}q_{\rho}g_{\nu\sigma}+q_{\mu}q_{\sigma}g_{\nu\rho}-q_{\nu}q_{\sigma}g_{\mu\rho}+q_{\nu}q_{\rho}g_{\mu\sigma})\\
+&\Pi_{b}^{P_A/P_{A_s}}\frac{1}{q^2}(-q_{\mu}q_{\rho}g_{\nu\sigma}+q_{\mu}q_{\sigma}g_{\nu\rho}\\
-&q_{\nu}q_{\sigma}g_{\mu\rho}+q_{\nu}q_{\rho}g_{\mu\sigma}),\\
\Pi_{\mu\nu}^{P_{B_k}/P_{B_{sk}}}(q^{2})=&\Pi_{(1)}^{P_{B_k}/P_{B_{sk}}}(q^2)\left(-g_{\mu\nu}+\frac{q_{\mu}q_{\nu}}{q^2}\right)\\
+&\Pi_{(0)}^{P_{B_k}/P_{B_{sk}}}(q^2)\left(\frac{q_{\mu}q_{\nu}}{q^2}\right),
\end{aligned}
\end{equation}
where $k=$1,2, $\Pi_{a}^{P_A/P_{A_s}}$ and $\Pi_{b}^{P_A/P_{A_s}}$ describes pure state contributions with quantum numbers $1^{++}$, $1^{-+}$ respectively, and $\Pi_{(1)}^{P_{B_k}/P_{B_{sk}}}$ and $\Pi_{(0)}^{P_{B_k}/P_{B_{sk}}}$ describes $1^{-+}$, $0^{++}$ respectively.
In mixing scenarios, we start from the off-diagonal mixed
correlator described in the previous sections, i.e.,
\begin{equation}
\begin{aligned}
\Pi^{M_{P_1}/M_{P_{s1}}}_{\mu\nu\sigma}(q^{2})&=\frac{i}{2}\int d^{4}x \,e^{iq\cdot x}  \langle 0|
T(j^{P_A/P_{A_s}}_{\mu\nu}(x)j^{P_{B_1}/P_{B_{s1}}+}_{\sigma}(0)\\
&+j^{P_{B_{1}}/P_{B_{s1}}}_{\sigma}(x)j^{P_A/P_{A_s}+}_{\mu\nu}(0) ) |0 \rangle, \\
\Pi^{M_{P_2}/M_{P_{s2}}}_{\mu\nu\sigma}(q^{2})&=\frac{i}{2}\int d^{4}x \,e^{iq\cdot x}  \langle 0|
T(j^{P_A/P_{A_s}}_{\mu\nu}(x)j^{P_{B_2}/P_{B_{s2}}+}_{\sigma}(0)\\
&+j^{P_{B_2}/P_{B_{s2}}}_{\sigma}(x)j^{P_A/P_{A_s}+}_{\mu\nu}(0) ) |0 \rangle,
\end{aligned}
\end{equation}
where $M_{P_k}/M_{P_{sk}}$, $k=$1,2, are mixed states assumed to result from the corresponding currents. The correlators have Lorentz structure
\begin{equation}
\begin{aligned}
&\Pi_{\mu\nu\sigma}^{M_{P_k}/M_{P_{sk}}}(q^{2})=\Pi^{M_{P_k}/M_{P_{sk}}}(q^{2})(-g_{\mu\sigma}q_{\nu}+g_{\nu\sigma}q_{\mu}).\\
\end{aligned}
\end{equation}

We follow same method used in previous sections. The mesonic structures, mass and coupling constant predictions, and related QCDSR parameters are collected in Table \ref{t5}.

\begin{table*}
\begin{ruledtabular}
\begin{center}
\caption{Summary of results for $1^{-+}$ molecular states. }
\label{t5} 
\begin{tabular}{cccccc}
State & Current structure & Mass/GeV & $\lambda$/$10^{-4}$GeV$^{10}$& $\sqrt{s_0}$/GeV & $\tau$ window/GeV$^{-2}$ \\
\hline
$P_A$& $J/\psi \rho$  & $4.658^{+0.05}_{-0.06}$ & $13.1^{+3.2}_{-3.1}$  & 5.15  & 0.24 -- 0.29 \\
$P_{A_s}$& $J/\psi f(980) $  & $4.694^{+0.05}_{-0.05}$ & $14.0^{+3.5}_{-2.9}$ & 5.2  & 0.24 -- 0.32 \\
$P_{B_1}$            & $D_0^*\bar{D}^*$  & $4.927^{+0.05}_{-0.04}$ & $34.1^{+7.7}_{-5.7}$ & 5.4 & 0.21 -- 0.30\\
$P_{B_2}$            & $D_1\bar{D}$  & $4.528^{+0.06}_{-0.05}$ & $9.7^{+2.9}_{-2.1}$ & 5.05 & 0.26 -- 0.31 \\
$P_{B_{s1}}$            & $D_{s0}^*\bar{D}_s^*$  & $4.999^{+0.04}_{-0.03}$ & $42.8^{+8.2}_{-6.8}$ & 5.5 & 0.21 -- 0.33\\
$P_{B_{s2}}$            & $D_{s1}\bar{D}_s$  & $4.642^{+0.05}_{-0.05}$ & $13.0^{+3.4}_{-2.8}$  & 5.15 & 0.25 -- 0.34\\
$M_{P_{1}}$   &  $J/\psi \rho$ - $D_0^*\bar{D}^*$ & $4.505^{+0.06}_{-0.04}$ & $0.401^{+0.096}_{-0.073}$ GeV$^{-1}$& 5.05 & 0.21 -- 0.30 \\
$M_{P_{2}}$   &  $J/\psi \rho$ - $D_1\bar{D}$ & $4.494^{+0.06}_{-0.06}$ & $0.240^{+0.060}_{-0.052}$ GeV$^{-1}$ & 5.05 & 0.23 -- 0.27 \\
$M_{P_{s1}}$   &  $J/\psi f(980)$ - $D_{s0}^*\bar{D}_s^*$ & $4.544^{+0.05}_{-0.05}$ & $0.405^{+0.085}_{-0.077}$ GeV$^{-1}$ & 5.1 & 0.21 -- 0.33\\
$M_{P_{s2}}$   &  $J/\psi f(980)$ - $D_{s1}\bar{D}_s$ & $4.536^{+0.06}_{-0.05}$ & $0.269^{+0.067}_{-0.055}$ GeV$^{-1}$ & 5.1 & 0.22 -- 0.31\\
\end{tabular}
\end{center}
\end{ruledtabular}
\end{table*}

All pure states have mass predictions over 4.5 GeV, and cannot be compatible with those known states which are probably $1^{-+}$ candidates ~\cite{Zyla:2020zbs}.

The mixing strength can then be estimated by computing the value of $N$. Note that the coupling constants of the mixed state correlator has the form
\begin{equation}
\begin{aligned}
\Pi^{M_{P_k}}_{\mu\nu\sigma}(q^{2})&=\frac{i}{2}\int d^{4}x \,e^{iq\cdot x}  \langle 0|
T(j^{P_A}_{\mu\nu}(x)j^{P_{B_k}+}_{\sigma}(0)\\
&+j^{P_{B_k}}_{\sigma}(x)j^{P_{A}+}_{\mu\nu}(0) ) |0 \rangle\\ &\sim \frac{\lambda_{P_A}}{M_H}(\varepsilon^{P_A}_{\mu}q_{\nu}-\varepsilon^{P_A}_{\nu}q_{\mu})(\lambda^*_{P_{B_k}}\varepsilon^{*P_{B_k}}_{\sigma})\\
&+\frac{\lambda^*_{P_A}}{M_H}(\varepsilon^{*P_A}_{\mu}q_{\nu}-\varepsilon^{*P_A}_{\nu}q_{\mu})(\lambda_{P_{B_k}}\varepsilon^{P_{B_k}}_{\sigma})+...\\
&=\frac{\lambda_{P_A}\lambda^*_{P_{B_k}}+\lambda^*_{P_A}\lambda_{P_{B_k}}}{M_H}(-g_{\mu\sigma}q_{\nu}+g_{\nu\sigma}q_{\mu})+...,
\end{aligned}
\end{equation}
where $k=1,2$, $\varepsilon^{P_A}$ and $\varepsilon^{P_{B_k}}$ are polarization vectors, and $ M_H$ represents the ground state mass of $P_A$. Analogues to $1^{++}$ channel, the values of $N$ obtained from Table \ref{t5} can be written as
\begin{equation}\label{firstn3}
\begin{aligned}
&N_{M_{P_1}}=\frac{0.401\textrm{GeV}^9\times M_{H}(4.658 \textrm{GeV})}{\sqrt{13.1}\textrm{GeV}^5\times\sqrt{34.1}\textrm{GeV}^5}=0.09\textrm{,}\\
&N_{M_{P_2}}=\frac{0.240\textrm{GeV}^9\times M_{H}(4.658 \textrm{GeV})}{\sqrt{13.1}\textrm{GeV}^5\times\sqrt{9.7}\textrm{GeV}^5}=0.10\textrm{,}\\
&N_{M_{P_{s1}}}=\frac{0.405\textrm{GeV}^9\times M_{H}(4.694 \textrm{GeV})}{\sqrt{14.0}\textrm{GeV}^5\times\sqrt{42.8}\textrm{GeV}^5}=0.08\textrm{,}\\
&N_{M_{P_{s2}}}=\frac{0.269\textrm{GeV}^9\times M_{H}(4.694 \textrm{GeV})}{\sqrt{14.0}\textrm{GeV}^5\times\sqrt{13.0}\textrm{GeV}^5}=0.09\textrm{,}
\end{aligned}
\end{equation}
where $M_H$ represents the corresponding $\bar{Q}Q\bar{q}q$ ground state mass. Like the $1^{--}$ channel, all the mixed states which have quantum numbers $1^{-+}$ are weakly mixed with corresponding currents, which becomes even more clear when we convert $N$ to $\widetilde{N}$,
\begin{equation}\label{ntilda3}
\begin{aligned}
&\widetilde{N}_{M_{P_1}}=\sin^2\left(\frac{\arcsin(0.09\times2)}{2}\right)=0.82\%,\\
&\widetilde{N}_{M_{P_2}}=\sin^2\left(\frac{\arcsin(0.10\times2)}{2}\right)=1.0\%,\\
&\widetilde{N}_{M_{P_{s1}}}=\sin^2\left(\frac{\arcsin(0.08\times2)}{2}\right)=0.64\%,\\
&\widetilde{N}_{M_{P_{s2}}}=\sin^2\left(\frac{\arcsin(0.09\times2)}{2}\right)=0.82\%.
\end{aligned}
\end{equation}
For the same reasons mentioned in the $1^{--}$ channel, $M_{P_1}$(4505) and $M_{P_{s1}}$(4494) are dominated by $\bar{Q}Q\bar{q}q$ components, $M_{P_2}$(4544) and $M_{P_{s2}}$(4536) are more likely dominated by $\bar{Q}qQ\bar{q}$ components.

 In Ref.~\cite{Wang:2016wwe}, authors have calculated states $P_{B_1}$ and $P_{B_2}$ with similar method and obtained the results:  mass $m_{P_{B_1}}=4.73^{+0.07}_{-0.07}$ with  $\sqrt{s_0}=5.2$ GeV$^{-2}$, and mass $m_{P_{B_2}}=4.60^{+0.08}_{-0.08}$ with  $\sqrt{s_0}=5.1$ GeV$^{-2}$, which are consistent with our results. The small difference of mass of $P_{B_1}$ is caused by different values of $\sqrt{s_0}$. Besides, authors in Ref.~\cite{Lee:2008gn} obtained mass of state $P_{B_1}$,  $m_{P_{B_1}}=4.19$ GeV. Our results are more supportive to the results in Ref.~\cite{Wang:2016wwe}.

\section{Non-perturbative effects of mixing strength}\label{sec:VI}
In fact, we can convert $\bar{Q}Q\bar{q}q$ and $\bar{Q}qQ\bar{q}$ states to each other through Fierz transformation. Generally,
\begin{equation}\label{fierz}
\begin{aligned}
(\bar{Q}\Gamma_1Q)(\bar{q}\Gamma_2q)&=\sum_{ijk} C_i(\bar{Q}\Gamma_jq)(\bar{q}\Gamma_kQ)\\
&+\sum_{lmn} C_l(\bar{Q}\Gamma_m\lambda^aq)(\bar{q}\Gamma_n\lambda^aQ),\\
(\bar{Q}\Gamma_1q)(\bar{q}\Gamma_2Q)&=\sum_{ijk} C_i(\bar{Q}\Gamma_jQ)(\bar{q}\Gamma_kq)\\
&+\sum_{lmn} C_l(\bar{Q}\Gamma_m\lambda^aQ)(\bar{q}\Gamma_n\lambda^aq),\\
\end{aligned}
\end{equation}
where $\Gamma_i$ are gamma matrices, $C_i$ are parameters corresponding to the related currents, and $\lambda^a$ are Gell-Mann matrices. That is, $\bar{Q}Q\bar{q}q$ currents can be decomposed into a series of $\bar{Q}qQ\bar{q}$  currents and a series of $\bar{Q}qQ\bar{q}$ color-octet currents, and vice versa. In this paper, we have computed two-point correlation functions of $\bar{Q}Q\bar{q}q$ and $\bar{Q}qQ\bar{q}$ currents. One can convert one current to a series of other kinds of currents, and make calculations analogous to a series of calculations of pure currents. For instance,
\begin{equation}\label{jajb}
\begin{aligned}
j^{Y_{A_1}}_{\mu}&=(\bar{c}c) (\bar{q}\gamma_{\mu}q)\\
&=\frac{-i}{2\sqrt{2}}\frac{i}{\sqrt{2}}[(\bar{c}\gamma_{\mu}q)(\bar{q}c)+(\bar{q}\gamma_{\mu}c)(\bar{c}q)]\\
&+\frac{-i}{2\sqrt{2}}\frac{i}{\sqrt{2}}[(\bar{c}\gamma_{\mu}\gamma_5q)(\bar{q}\gamma_5c)-(\bar{q}\gamma_{\mu}\gamma_5c)(\bar{c}\gamma_5q)]\\
&+...\\
&=\frac{-i}{2\sqrt{2}}j_{\mu}^{Y_{B_1}}+\frac{-i}{2\sqrt{2}}j_{\mu}^{Y_{B_2}}+...,
\end{aligned}
\end{equation}
where $j^{Y_{A_1}}_{\mu}$, $j_{\mu}^{Y_{B_1/B_2}}$ are defined in Section \ref{sec:IV}. When we compute two-point correlation functions of $j_{\mu}^{Y_{A_1}}$ and  $j_{\mu}^{Y_{B_1/B_2}}$, it seems that the result may highlight states $Y_{B_1}$/$Y_{B_2}$, and the parameters of the current decomposition are likely to be related directly to mixing strength. However, our calculations show different results. Although the contributions in perturbative terms from different currents 
(e.g., $Y_{B_1}$ and $Y_{B_2}$) will be suppressed, QCDSR calculations are sensitive to the changes of borel window and threshold $s_0$, which depend on contributions of non-perturbative terms. Moreover, the mixing strength is related to both decay constants and parameters of the corresponding currents from the Fierz transformation, and the decay constants are also sensitive to the Borel window, which again depend on non-perturbative terms. 
To clarify this we have computed another two $\bar{Q}Q\bar{q}q$ and $\bar{Q}qQ\bar{q}$ currents and their mixed state,
\begin{equation}
\begin{aligned}
j^{Z_A}_{\mu}(x)=&\bar{c}(x)\gamma_{\mu}c(x) \bar{q}(x)\gamma_{5}q(x)\\
j^{Z_{B}}_{\mu\nu}(x)=&\frac{i}{\sqrt{2}}[\bar{c}(x)\gamma_{\mu}q(x)\bar{q}(x)\gamma_{\nu}c(x)\\
-&\bar{q}(x)\gamma_{\mu}c(x)\bar{c}(x)\gamma_{\nu}q(x)],
\end{aligned}
\end{equation}
where $Z$ denotes the $1^{+-}$ state, and the subscript $A$ of $Z$ represents the $\bar{Q}Q\bar{q}q$ scenario while $B$ represents the $\bar{Q}qQ\bar{q}$ scenario. The mixed state is described by
\begin{equation}
\begin{aligned}
\Pi^{M_{Z}}_{\mu\nu\sigma}(q^{2})&=\frac{i}{2}\int d^{4}x \,e^{iq\cdot x}  \langle 0|
T(j^{Z_A}_{\sigma}(x)j^{Z_{B}+}_{\mu\nu}(0)\\
&+j^{Z_{B}}_{\mu\nu}(x)j^{P_A+}_{\sigma}(0) ) |0 \rangle, \\
\end{aligned}
\end{equation}
where $M_{Z}$ is assumed mixing from corresponding currents. The hadronic structures along with results of mass, coupling constant and mixing strength predictions are collected in Table \ref{t7}. 
\begin{table*}
\begin{ruledtabular}
\begin{center}
\caption{Summary of results for $1^{+-}$  states.}
\label{t7} 
\begin{tabular}{cccccc}
State & Current structure & Mass/GeV & $\lambda$/$10^{-4}$GeV$^{10}$& $\sqrt{s_0}$/GeV & $\tau$ window/GeV$^{-2}$ \\
\hline
$Z_A$& $J/\psi \eta$  & $3.578^{+0.08}_{-0.08}$ & $1.04^{+0.36}_{-0.27}$  & 4.2  & 0.33 -- 0.34 \\
$Z_{B}$            & $D^*\bar{D}^*$  & $4.018^{+0.06}_{-0.06}$ & $2.90^{+0.88}_{-0.67}$ & 4.55 & 0.29 -- 0.36\\
$M_{Z}$   &  $J/\psi \eta$ - $D^*\bar{D}^*$ & $3.563^{+0.07}_{-0.06}$ & $0.054^{+0.017}_{-0.012}$ GeV$^{-1}$& 4.1 & 0.31 -- 0.35 \\
\end{tabular}
\end{center}
\end{ruledtabular}
\end{table*}

Compared to $1^{++}$ currents $j_{\mu}^{X_{A/B_1}}$ and their mixed two-point correlator $\Pi^{M_{X_1}}_{\mu\nu\sigma}(q^{2})$, which are given in Eq.~(\ref{++}) and Eq.~(\ref{++2}), $j_{\mu}^{Z_{A/B}}$ and $\Pi^{M_{Z}}_{\mu\nu\sigma}(q^{2})$  have similar structures.  Due to our previous calculations in Section \ref{sec:II}, $M_{X_1}$ is relatively strongly mixed with different components, and $M_Z$ is supposed to have similar properties. However, the resulting mixing strength of $M_Z$ is
\begin{equation}\label{firstn8}
\begin{aligned}
&N_{M_{Z}}=\frac{0.054\textrm{GeV}^9\times M_{H}(4.018 \textrm{GeV})}{\sqrt{1.04}\textrm{GeV}^5\times\sqrt{2.90}\textrm{GeV}^5}=0.125\textrm{,}\\
&\widetilde{N}_{M_{Z}}=\sin^2(\frac{\arcsin(0.125\times2)}{2})=1.6\%.\\
\end{aligned}
\end{equation}
Compared to $M_{X_1}$($N_{M_{X_1}}$=0.349, $\widetilde{N}_{M_{X_1}}$=14\%), the mass predictions of two parts of mixed state $M_Z$ differ, and although the contributions of perturbative terms in two-point correlator functions are similar, the mixing strength of two states are quite different. Hence we suggest that mixing strength is much sensitive to the Borel window, threshold $s_0$, mass prediction, and decay constant, which are all influenced by non-perturbative terms in QCDSR calculations.
\section{Summary}
In this paper  we  used QCD sum-rules to calculate the mass spectrum of $\bar{Q}Q\bar{q}q$ and $\bar{Q}qQ\bar{q}$  states.
Such  states strongly couple to   $\bar{Q}Q\bar{q}q$ or $\bar{Q}qQ\bar{q}$  currents. So state's components of 
 $\bar{Q}Q\bar{q}q$ and $\bar{Q}qQ\bar{q}$  can be mixed with each other. Such mixing can be studied via 
the mixed correlators of  $\bar{Q}Q\bar{q}q$ and $\bar{Q}qQ\bar{q}$ currents.  Our studies focus on mixing strength which may 
determine  whether the mixing picture accommodates candidates  which have more than single dominant decay modes.   

\begin{table*}
\begin{ruledtabular}
\begin{center}
\caption{Summary of mixed state results.  }
\label{t6} 
\begin{tabular}{cccccc}
Mixed state & Mass/GeV & $N$ & $\widetilde{N}$ & Dominant part&  Possible Candidate \\
\hline
$M_{X_1}$   & $3.987^{+0.06}_{-0.06}$  & 0.349 & 14\% & $\bar{Q}qQ\bar{q}$ & $X(3872)$ \\$M_{X_2}$   & $4.945^{+0.08}_{-0.06}$  & 0.285 & 9.0\% &$\bar{Q}qQ\bar{q}$ & -\\$M_C$   & $3.818^{+0.03}_{-0.02}$  & 0.125 & 1.6\% & $\bar{Q}Q$ & $X(3872)$ \\$M_{Y_1}$   &  $4.770^{+0.07}_{-0.06}$ & 0.15 & 2.3\% & $\bar{Q}qQ\bar{q}$ & -\\$M_{Y_2}$   &  $4.266^{+0.08}_{-0.08}$  & 0.11 & 1.2\% & $\bar{Q}Q\bar{q}q$ & $Y(4230)$\\$M_{Y_{s1}}$   & $4.610^{+0.05}_{-0.06}$ & 0.06 & $<$1\% & $\bar{Q}Q\bar{q}q$ & $Y(4660)$\\$M_{Y_{s2}}$   & $4.450^{+0.05}_{-0.06}$  & 0.11 & 1.2\% & $\bar{Q}qQ\bar{q}$ & $Y(4415)$\\$M_{P_1}$   &  $4.505^{+0.06}_{-0.04}$  & 0.09 & $<$1\% & $\bar{Q}Q\bar{q}q$ & -\\$M_{P_2}$   & $4.494^{+0.06}_{-0.06}$  & 0.10 & 1.0\% & $\bar{Q}qQ\bar{q}$ &-\\$M_{P_{s1}}$   & $4.544^{+0.05}_{-0.05}$ & 0.08 & $<$1\% & $\bar{Q}Q\bar{q}q$ & -\\$M_{P_{s2}}$   &  $4.536^{+0.06}_{-0.05}$  & 0.09 & $<$1\% & $\bar{Q}qQ\bar{q}$ & -\\
\end{tabular}
\end{center}
\end{ruledtabular}
\end{table*}
We list all the  mixed states results in Table \ref{t6}. The uncertainties of masses are less than 5\%, and the uncertainties of coupling constants are about 25\%, which are induced by uncertainties of input parameters and threshold $s_0$. The relations $\sqrt{s_0}=M_H+\Delta_s$ and 40\%-10\%   are required to determine  the window of  $\tau$.
 These two conditions are not always satisfied well. In some cases, the windows of $\tau$  is very narrow.  If higher dimension condensates are considered, we may reconsider the constraint of 40\%-10\% and the situation may change. 

For the $1^{++}$ channel, we find that two states $M_{X_1}$(3987) and $M_{X_2}$(4945) are relatively strongly mixed with $\bar{Q}Q\bar{q}q$ and $\bar{Q}qQ\bar{q}$ components. Furthermore, we estimate the ratio of decay width of two kinds of decay modes of $M_{X_1}$(3987), which is roughly consistent with experimental results for the  $X(3872)$. When we consider the mixing state combined with $\bar{Q}Q$ and $\bar{Q}qQ\bar{q}$,  we revisit the result in Ref.\,\cite{Matheus:2009vq} with the new technique in Ref.\,\cite{Chen:2019buw}. The result argues that $\bar{Q}Q$ is the dominant part of $X(3872)$,  which can explain the latest observation to $X(3872)$ of LHCb\cite{LHCb:2020sey}. Our calculations just support these two components  can relatively strongly mix with each other in quantum numbers $1^{++}$. 

In other quantum number channels, states are found to be weakly mixed. However, the calculations of these states is still meaningful to help us establish the physical structure of corresponding state. For instance, pure $\bar{Q}Q\bar{q}q$  $Y_{A_1}$(4207) and $Y_{A_{s2}}$(4610) configurations are good candidates for $Y(4230)$ and $Y(4660)$ respectively. However, by checking assumed mixed states mixing with $\bar{Q}Q\bar{q}q$ and $\bar{Q}qQ\bar{q}$ molecular states, we find these candidates have small components  of $\bar{Q}qQ\bar{q}$ which is inconsistent 
with the fact that  $Y(4230)$ and $Y(4660)$ have  more abundant decay modes. For the same reasons, we can establish the dominant part of $Y(4415)$. Our result suggests that $Y(4415)$ is dominated by $\bar{Q}qQ\bar{q}$, and agrees with the absence of $K$ meson in observed decay final states. But $Y(4415)$ still has a small component of $\bar{Q}Q\bar{q}q$.  These states  may therefore  have more a complicated construction, for instance, 
$\bar{q} q$ could be a color-octet state. 
Other models, such as the tetraquark model,  maybe valuable. By using the Fierz transformations, tetraquark currents can be decomposed into various molecular currents and color-octet currents, to show  more mixed effects of different possible states~\cite{Nielsen:2009uh}.  Since the mixing effects are normally small, the studies via tetraquark currents cannot distinguish details of the mixing between the different currents and only give the average of those currents. So the  tetraquark model is not a self-verifying because it cannot 
show which parts (via the Fierz transformations) interact with each other strongly and others do not. 
It  may also meet challenges for quantitative descriptions of XYZ states.

The calculations based on pure  molecular currents have been criticized because there is a large background of two free mesons spectrum. If the states indeed have an  absolutely  dominate decay mode\cite{Guo:2017jvc}, there is no problem(actually, the mass of molecule state is close to that of two free mesons). Otherwise, the mixing pattern must be taken into account. The mixing  of  typical molecular currents  $\bar{Q}Q\bar{q}q$ and  $\bar{Q}qQ\bar{q}$  are suppressed (perturbatively) by small coefficients of Fierz transformations, so the background of two free mesons spectrum also 
is suppressed.  Non-perturbative corrections play more important roles in the mixing correlator, which can distinguish the real four-quark resonance from two free mesons.  It should be the essential feature of the mixing pattern.   Our calculations show that the mixing pattern is consistent with some of  XYZ states, but  fail in many others.  Since the mixing correlator is normalized by two diagonal correlators which may be affected by a large background of two free meson spectrum,  the real mixture  may be larger than our estimate. How to remove 
the background of two free meson spectrum is still a big problem. 

\begin{acknowledgments}
This work is supported by NSFC (under grants 11175153 and 11205093) and by the Natural Sciences and Engineering Research Council of Canada (NSERC).
\end{acknowledgments}

\appendix
\section{QCDSR analysis results}\label{Appendix:A}
Here we show the $\tau$ dependence of $M_H^2$ defined in Eq.~(\ref{2.10}) for all mixed states.


\begin{figure*}[htbp]
\centering
\subfigure[$M_H^2$ 
dependence on
$\tau$ for  $M_{X_1}$. The solid line represents $\sqrt{s_0}=4.4$GeV, and the dashed line and the dotted lines 
respectively represent 
$\sqrt{s_0}=4.4\pm0.1$GeV .]{ \label{fig1} 
\includegraphics[width=0.9\columnwidth]{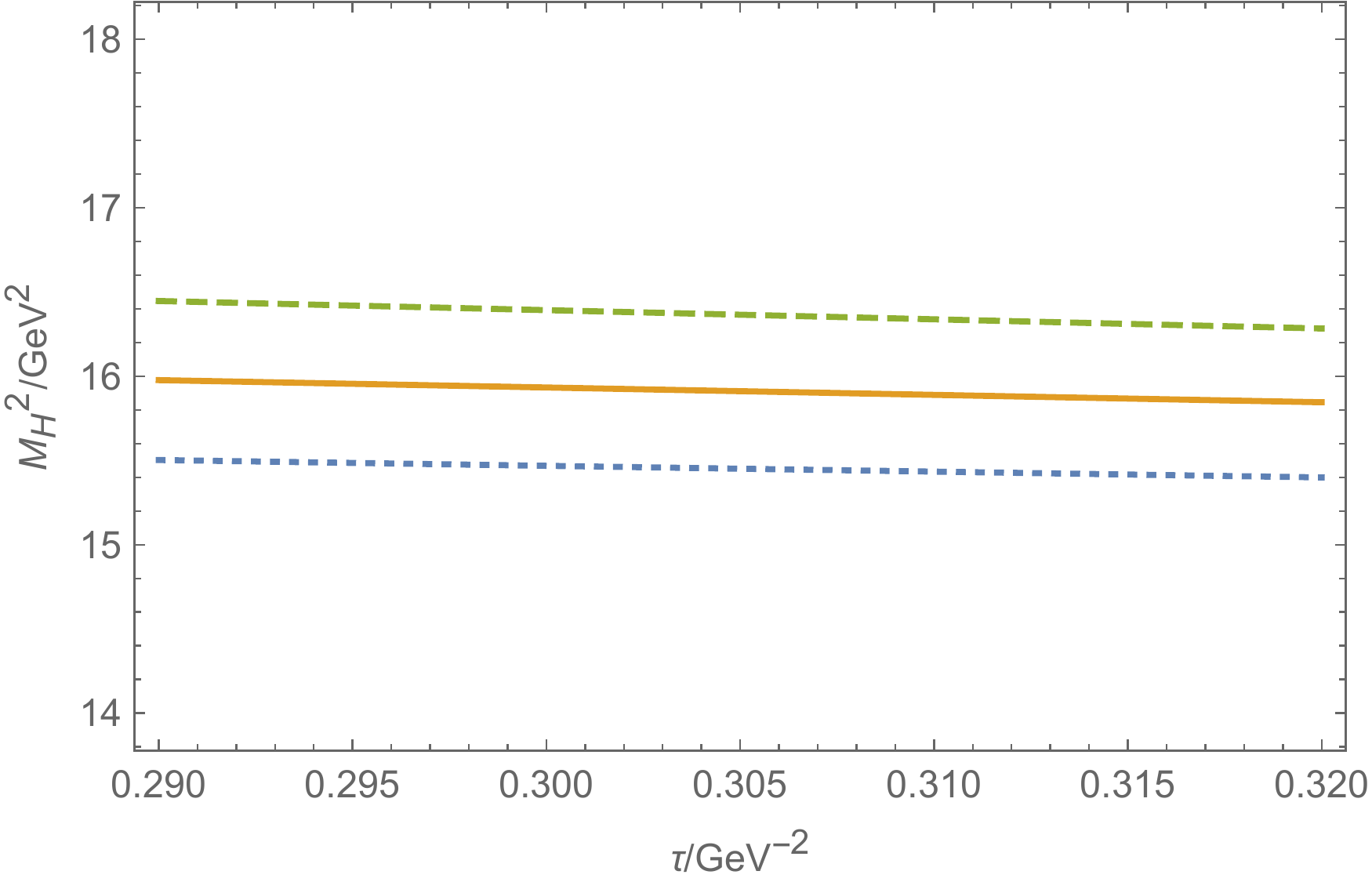}}
\hspace{1cm}
\subfigure[$M_H^2$  dependence on $\tau$ for $M_{X_2}$. The solid line 
represents
$\sqrt{s_0}=5.45$GeV, and the dashed line and the dotted lines 
respectively represent
$\sqrt{s_0}=5.45\pm0.1$GeV.]{ \label{fig2} 
\includegraphics[width=0.9\columnwidth]{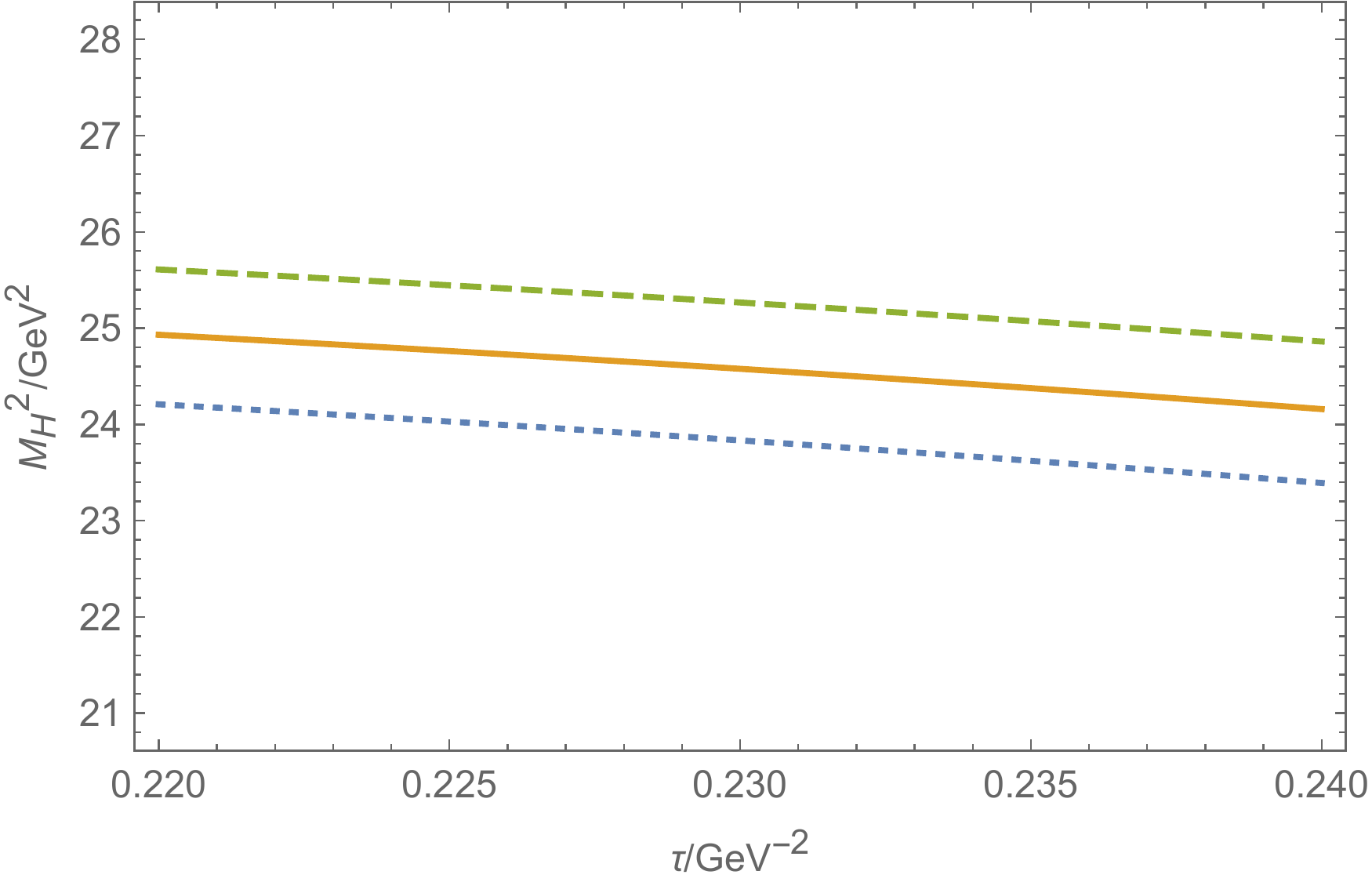}}
\caption{$M_H^2$ behaviors on $\tau$ for $1^{++}$ mixed states.}
\end{figure*}

\begin{figure*}[htbp]
\centering
\subfigure[$M_H^2$ 
dependence on $\tau$ for $M_{Y_1}$. The solid line 
represents
$\sqrt{s_0}=5.3$GeV, and the dashed line and the dotted line 
respectively represent
$\sqrt{s_0}=5.3\pm0.1$GeV.]{ \label{fig3} 
\includegraphics[width=0.9\columnwidth]{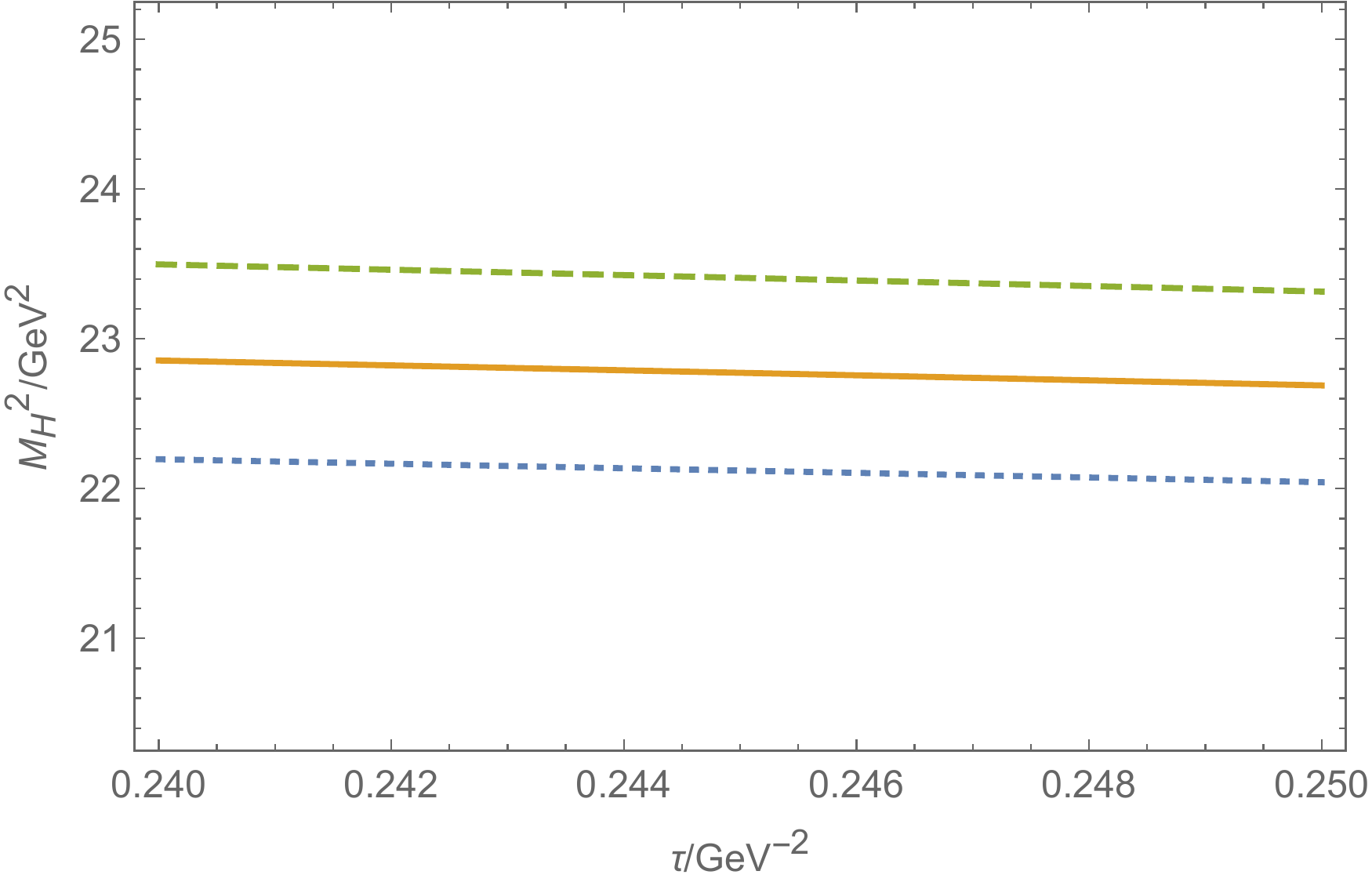}}
\hspace{1cm}
\subfigure[ $M_H^2$ 
dependence on
$\tau$ for $M_{Y_2}$. The solid line 
represents
$\sqrt{s_0}=4.95$GeV, and the dashed line and the dotted line 
respectively represent
$\sqrt{s_0}=4.95\pm0.1$GeV.]{ \label{fig4} 
\includegraphics[width=0.9\columnwidth]{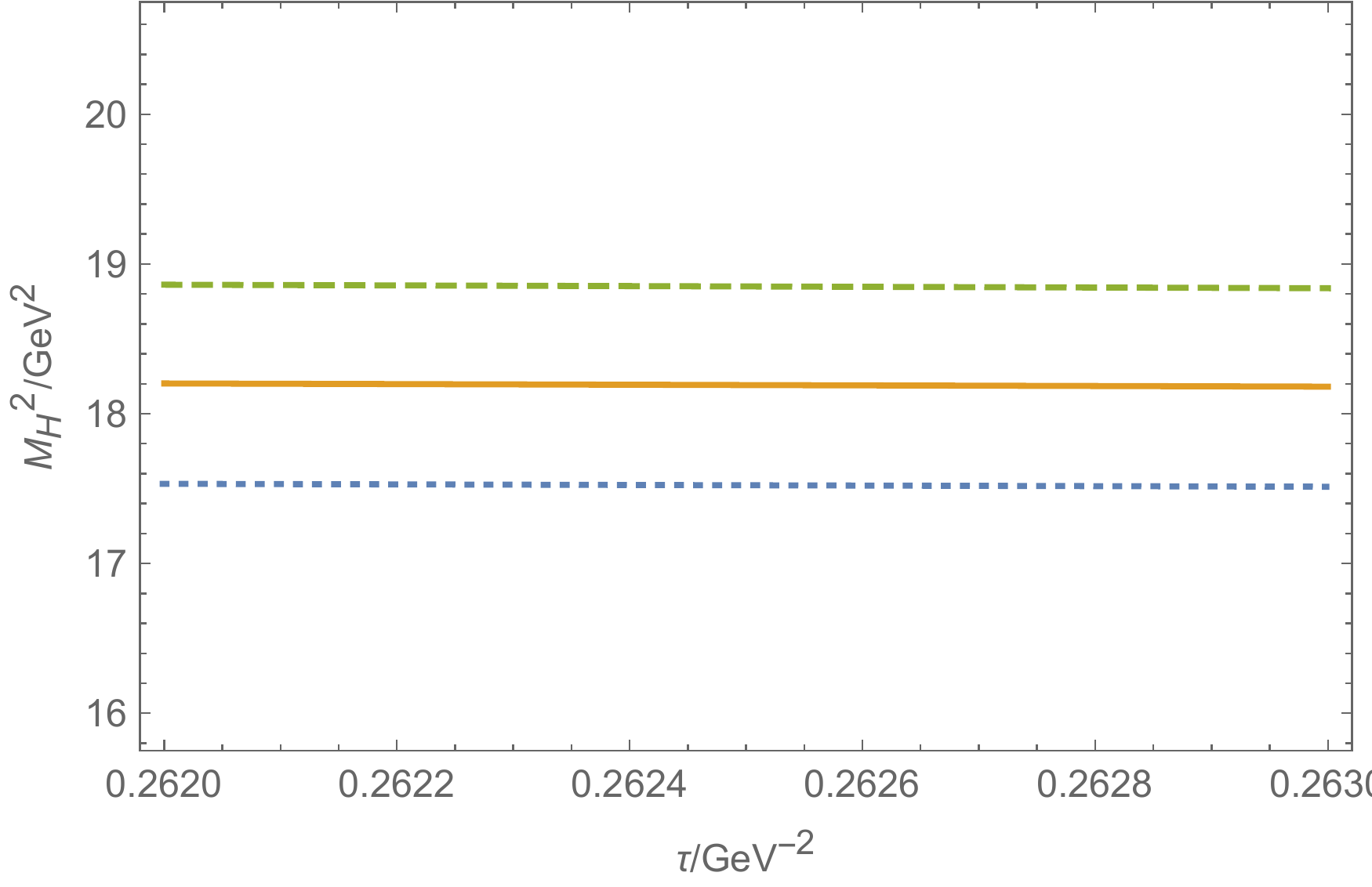}}
\\
\subfigure[$M_H^2$ 
dependence on 
$\tau$ 
for $M_{Y_{s1}}$. The solid line 
represents
$\sqrt{s_0}=5.1$GeV, and the dashed line and the dotted line 
respectively represent
$\sqrt{s_0}=5.1\pm0.1$GeV.]{ \label{fig5} 
\includegraphics[width=0.9\columnwidth]{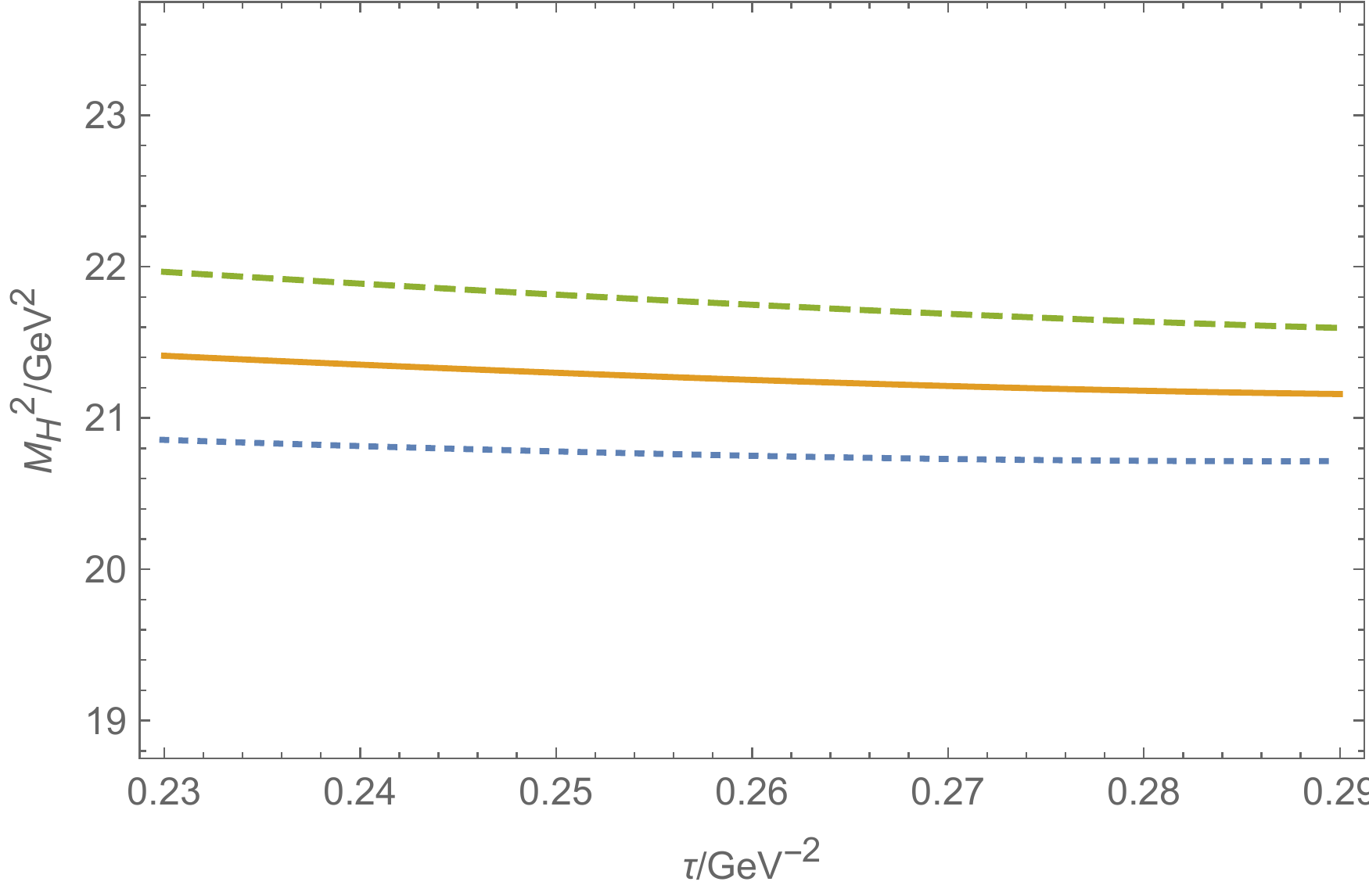}}
\hspace{1cm}
\subfigure[$M_H^2$ 
dependence on
$\tau$ for $M_{Y_{s2}}$. The solid line represents
$\sqrt{s_0}=4.95$GeV, and the dashed line and the dotted line 
respectively represent
$\sqrt{s_0}=4.95\pm0.1$GeV.]{ \label{fig6} 
\includegraphics[width=0.9\columnwidth]{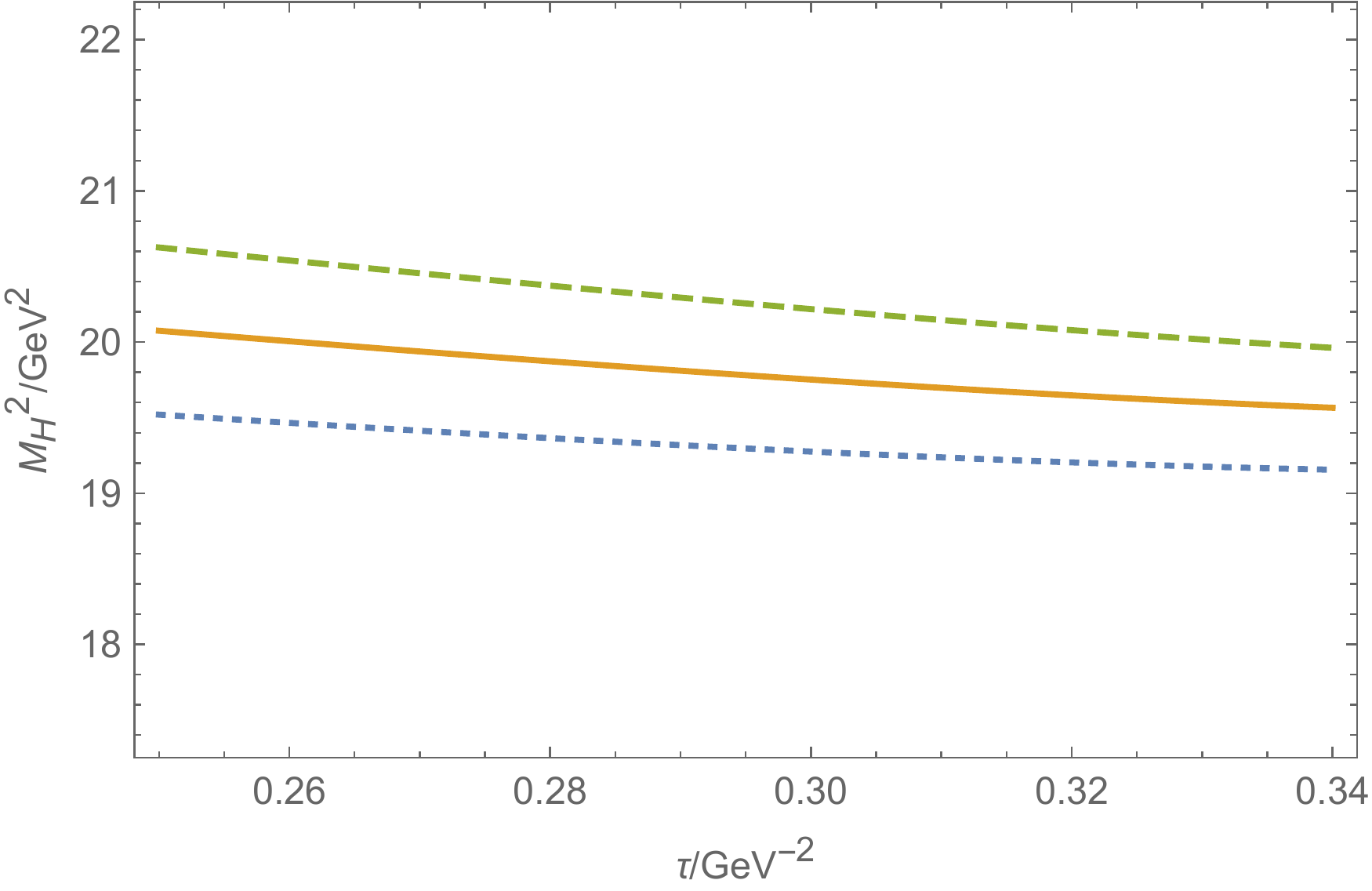}}
\caption{$M_H^2$ behaviors on$\tau$ for $1^{--}$ mixed states.}
\end{figure*}

\begin{figure*}[htbp]
\centering
\subfigure[$M_H^2$ 
dependence on
$\tau$ for $M_{P_1}$. The solid line 
represents
$\sqrt{s_0}=5.05$GeV, and the dashed line and the dotted line 
respectively represent
$\sqrt{s_0}=5.05\pm0.1$GeV.]{ \label{fig7} 
\includegraphics[width=0.9\columnwidth]{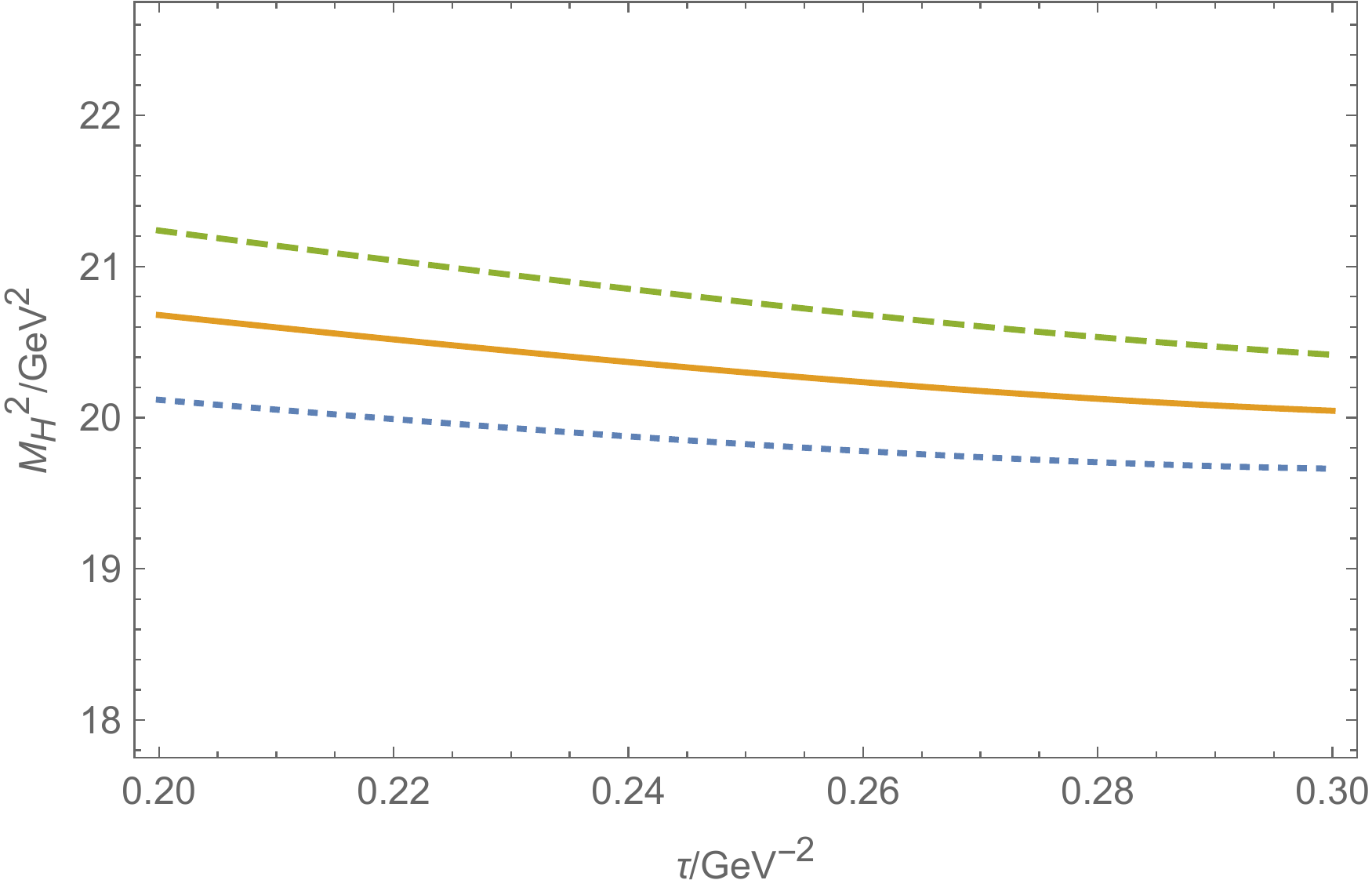}}
\hspace{1cm}
\subfigure[$M_H^2$ 
dependence on 
$\tau$ for $M_{P_2}$. The solid line 
represents
$\sqrt{s_0}=5.05$GeV, and the dashed line and the dotted line 
respectively represent 
$\sqrt{s_0}=5.05\pm0.1$GeV.]{ \label{fig8} 
\includegraphics[width=0.9\columnwidth]{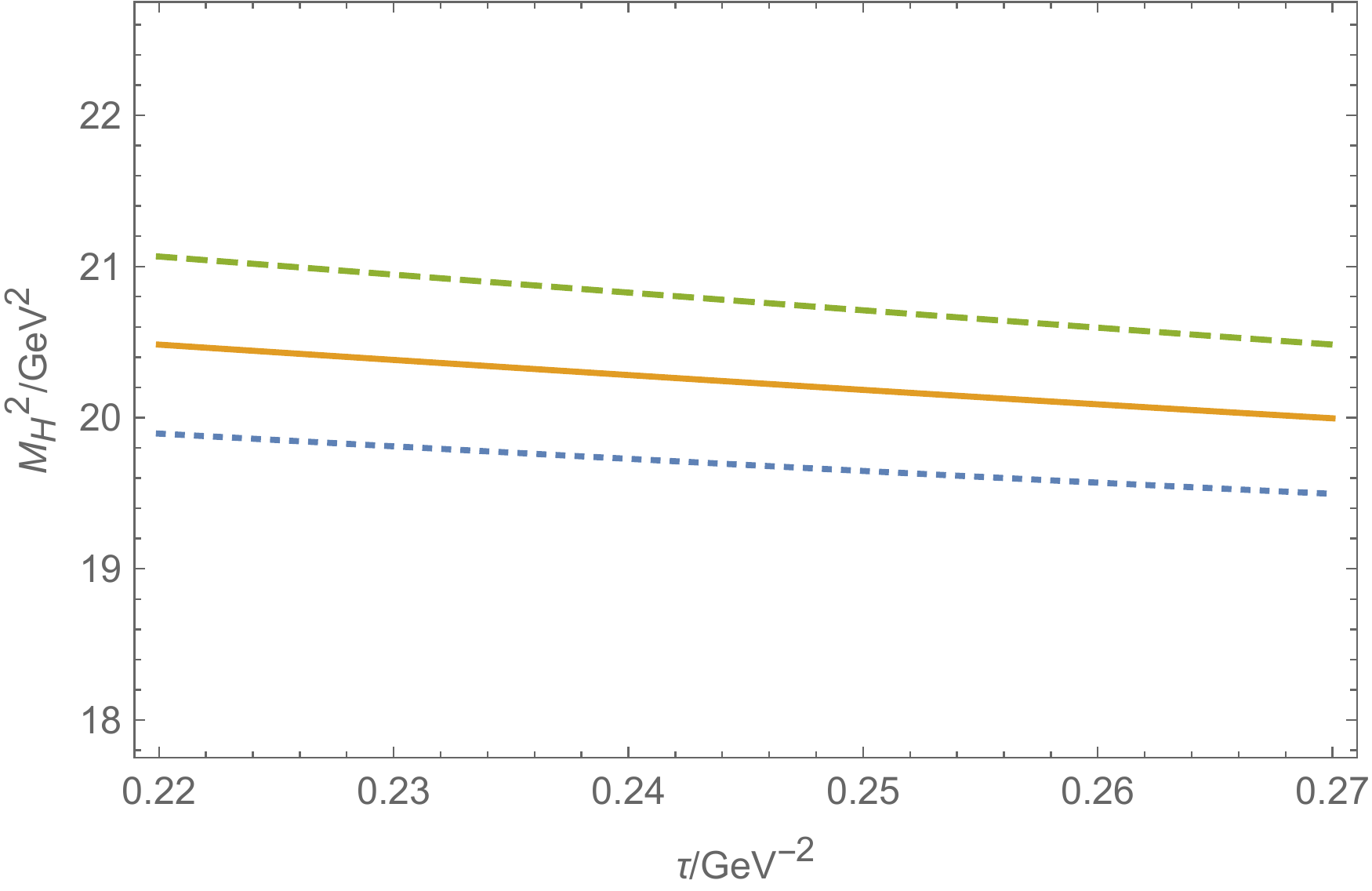}}
\\
\subfigure[ $M_H^2$ 
dependence on
$\tau$ for $M_{P_{s1}}$. The solid line represents
$\sqrt{s_0}=5.1$GeV, and the dashed line and the dotted line 
respectively represent 
$\sqrt{s_0}=5.1\pm0.1$GeV.]{ \label{fig9} 
\includegraphics[width=0.9\columnwidth]{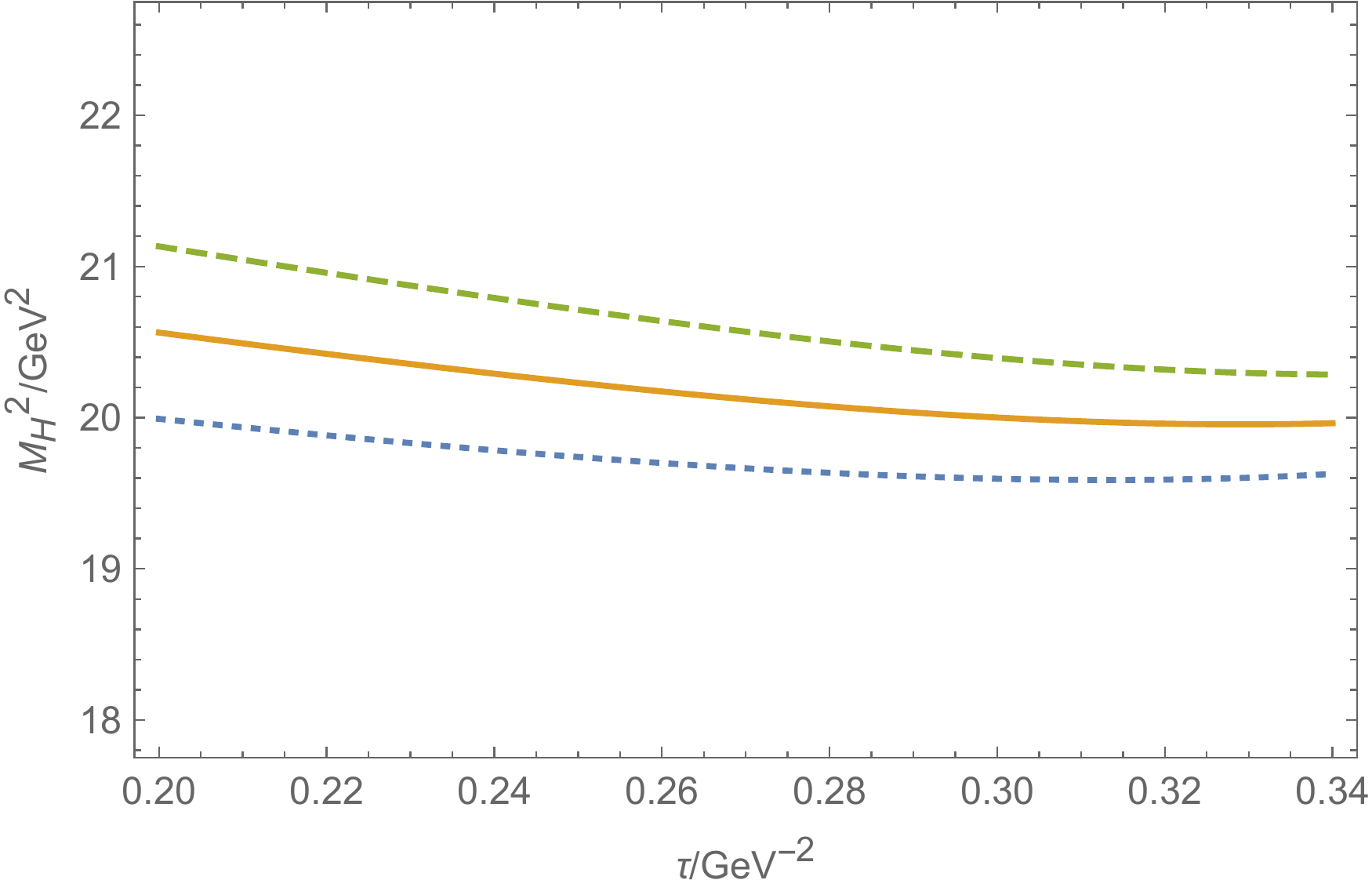}}
\hspace{1cm}
\subfigure[ $M_H^2$ 
dependence on 
$\tau$ for $M_{P_{s2}}$. The solid line represents
$\sqrt{s_0}=5.1$GeV, and the dashed line and the dotted line 
respectively represent 
$\sqrt{s_0}=5.1\pm0.1$GeV.]{ \label{fig10} 
\includegraphics[width=0.9\columnwidth]{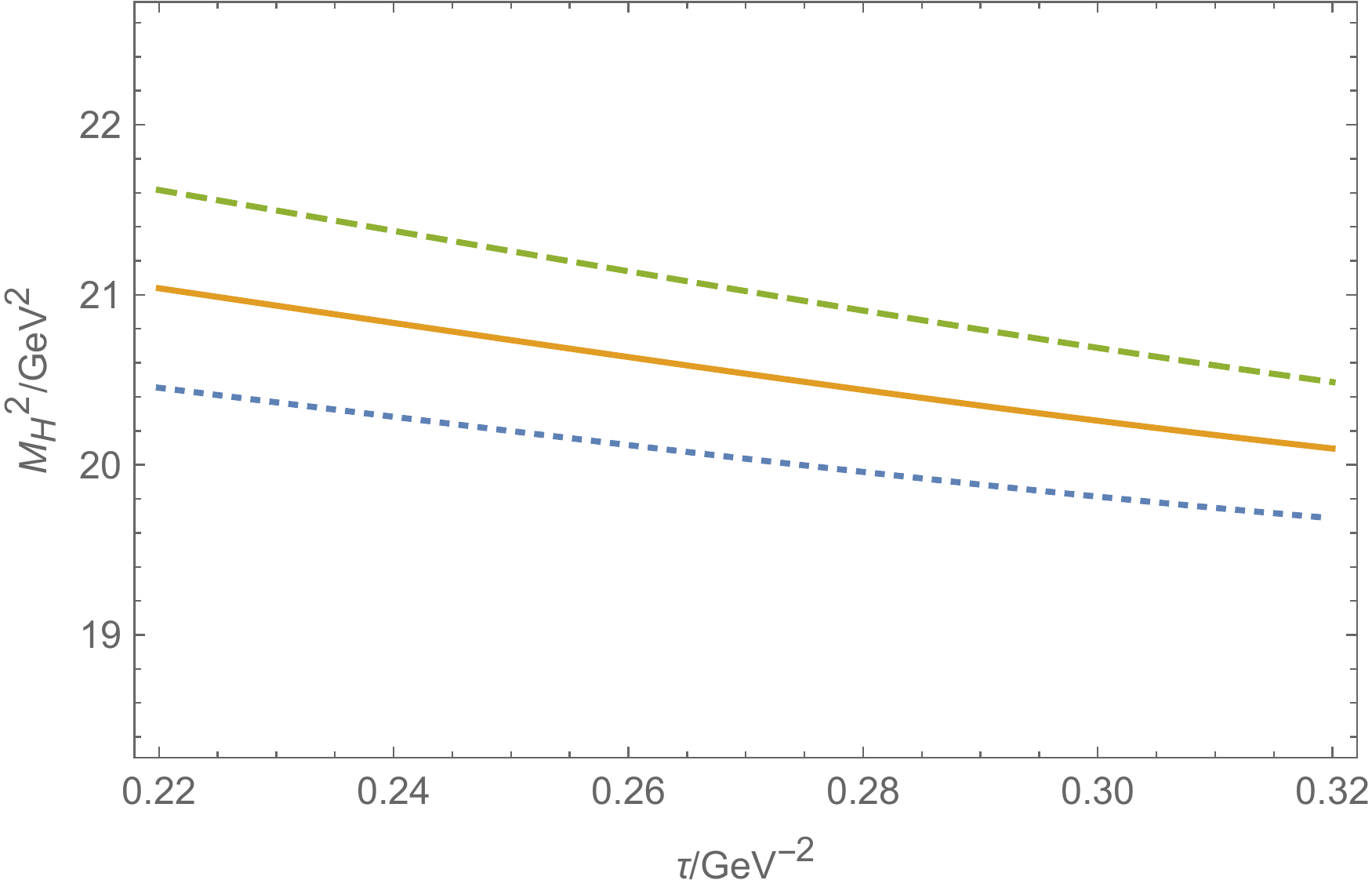}}
\caption{$M_H^2$ behaviors on $\tau$ for $1^{-+}$ mixed states.}
\end{figure*}
\nocite{*}

\bibliography{apssamp}

\providecommand{\noopsort}[1]{}\providecommand{\singleletter}[1]{#1}%
\begin{thebibliography}{45}%
\makeatletter
\providecommand \@ifxundefined [1]{%
 \@ifx{#1\undefined}
}%
\providecommand \@ifnum [1]{%
 \ifnum #1\expandafter \@firstoftwo
 \else \expandafter \@secondoftwo
 \fi
}%
\providecommand \@ifx [1]{%
 \ifx #1\expandafter \@firstoftwo
 \else \expandafter \@secondoftwo
 \fi
}%
\providecommand \natexlab [1]{#1}%
\providecommand \enquote  [1]{``#1''}%
\providecommand \bibnamefont  [1]{#1}%
\providecommand \bibfnamefont [1]{#1}%
\providecommand \citenamefont [1]{#1}%
\providecommand \href@noop [0]{\@secondoftwo}%
\providecommand \href [0]{\begingroup \@sanitize@url \@href}%
\providecommand \@href[1]{\@@startlink{#1}\@@href}%
\providecommand \@@href[1]{\endgroup#1\@@endlink}%
\providecommand \@sanitize@url [0]{\catcode `\\12\catcode `\$12\catcode
  `\&12\catcode `\#12\catcode `\^12\catcode `\_12\catcode `\%12\relax}%
\providecommand \@@startlink[1]{}%
\providecommand \@@endlink[0]{}%
\providecommand \url  [0]{\begingroup\@sanitize@url \@url }%
\providecommand \@url [1]{\endgroup\@href {#1}{\urlprefix }}%
\providecommand \urlprefix  [0]{URL }%
\providecommand \Eprint [0]{\href }%
\providecommand \doibase [0]{https://doi.org/}%
\providecommand \selectlanguage [0]{\@gobble}%
\providecommand \bibinfo  [0]{\@secondoftwo}%
\providecommand \bibfield  [0]{\@secondoftwo}%
\providecommand \translation [1]{[#1]}%
\providecommand \BibitemOpen [0]{}%
\providecommand \bibitemStop [0]{}%
\providecommand \bibitemNoStop [0]{.\EOS\space}%
\providecommand \EOS [0]{\spacefactor3000\relax}%
\providecommand \BibitemShut  [1]{\csname bibitem#1\endcsname}%
\let\auto@bib@innerbib\@empty
\bibitem [{\citenamefont {Brambilla}\ \emph {et~al.}(2020)\citenamefont
  {Brambilla}, \citenamefont {Eidelman}, \citenamefont {Hanhart}, \citenamefont
  {Nefediev}, \citenamefont {Shen}, \citenamefont {Thomas}, \citenamefont
  {Vairo},\ and\ \citenamefont {Yuan}}]{Brambilla:2019esw}%
  \BibitemOpen
  \bibfield  {author} {\bibinfo {author} {\bibfnamefont {N.}~\bibnamefont
  {Brambilla}}, \bibinfo {author} {\bibfnamefont {S.}~\bibnamefont {Eidelman}},
  \bibinfo {author} {\bibfnamefont {C.}~\bibnamefont {Hanhart}}, \bibinfo
  {author} {\bibfnamefont {A.}~\bibnamefont {Nefediev}}, \bibinfo {author}
  {\bibfnamefont {C.-P.}\ \bibnamefont {Shen}}, \bibinfo {author}
  {\bibfnamefont {C.~E.}\ \bibnamefont {Thomas}}, \bibinfo {author}
  {\bibfnamefont {A.}~\bibnamefont {Vairo}},\ and\ \bibinfo {author}
  {\bibfnamefont {C.-Z.}\ \bibnamefont {Yuan}},\ }\bibfield  {title} {\bibinfo
  {title} {{The $XYZ$ states: experimental and theoretical status and
  perspectives}},\ }\href {https://doi.org/10.1016/j.physrep.2020.05.001}
  {\bibfield  {journal} {\bibinfo  {journal} {Phys. Rept.}\ }\textbf {\bibinfo
  {volume} {873}},\ \bibinfo {pages} {1} (\bibinfo {year} {2020})},\ \Eprint
  {https://arxiv.org/abs/1907.07583} {arXiv:1907.07583 [hep-ex]} \BibitemShut
  {NoStop}%
\bibitem [{\citenamefont {Zyla}\ \emph {et~al.}(2020)\citenamefont {Zyla} \emph
  {et~al.}}]{Zyla:2020zbs}%
  \BibitemOpen
  \bibfield  {author} {\bibinfo {author} {\bibfnamefont {P.~A.}\ \bibnamefont
  {Zyla}} \emph {et~al.} (\bibinfo {collaboration} {Particle Data Group}),\
  }\bibfield  {title} {\bibinfo {title} {{Review of Particle Physics}},\ }\href
  {https://doi.org/10.1093/ptep/ptaa104} {\bibfield  {journal} {\bibinfo
  {journal} {PTEP}\ }\textbf {\bibinfo {volume} {2020}},\ \bibinfo {pages}
  {083C01} (\bibinfo {year} {2020})}\BibitemShut {NoStop}%
\bibitem [{\citenamefont {Albuquerque}\ \emph {et~al.}(2019)\citenamefont
  {Albuquerque}, \citenamefont {Dias}, \citenamefont {Khemchandani},
  \citenamefont {Mart\'\i{}nez~Torres}, \citenamefont {Navarra}, \citenamefont
  {Nielsen},\ and\ \citenamefont {Zanetti}}]{Albuquerque:2018jkn}%
  \BibitemOpen
  \bibfield  {author} {\bibinfo {author} {\bibfnamefont {R.~M.}\ \bibnamefont
  {Albuquerque}}, \bibinfo {author} {\bibfnamefont {J.~M.}\ \bibnamefont
  {Dias}}, \bibinfo {author} {\bibfnamefont {K.~P.}\ \bibnamefont
  {Khemchandani}}, \bibinfo {author} {\bibfnamefont {A.}~\bibnamefont
  {Mart\'\i{}nez~Torres}}, \bibinfo {author} {\bibfnamefont {F.~S.}\
  \bibnamefont {Navarra}}, \bibinfo {author} {\bibfnamefont {M.}~\bibnamefont
  {Nielsen}},\ and\ \bibinfo {author} {\bibfnamefont {C.~M.}\ \bibnamefont
  {Zanetti}},\ }\bibfield  {title} {\bibinfo {title} {{QCD sum rules approach
  to the $X,~Y$ and $Z$ states}},\ }\href
  {https://doi.org/10.1088/1361-6471/ab2678} {\bibfield  {journal} {\bibinfo
  {journal} {J. Phys. G}\ }\textbf {\bibinfo {volume} {46}},\ \bibinfo {pages}
  {093002} (\bibinfo {year} {2019})},\ \Eprint
  {https://arxiv.org/abs/1812.08207} {arXiv:1812.08207 [hep-ph]} \BibitemShut
  {NoStop}%
\bibitem [{\citenamefont {Chen}\ and\ \citenamefont {Zhu}(2011)}]{Chen:2010ze}%
  \BibitemOpen
  \bibfield  {author} {\bibinfo {author} {\bibfnamefont {W.}~\bibnamefont
  {Chen}}\ and\ \bibinfo {author} {\bibfnamefont {S.-L.}\ \bibnamefont {Zhu}},\
  }\bibfield  {title} {\bibinfo {title} {{The Vector and Axial-Vector
  Charmonium-like States}},\ }\href
  {https://doi.org/10.1103/PhysRevD.83.034010} {\bibfield  {journal} {\bibinfo
  {journal} {Phys. Rev. D}\ }\textbf {\bibinfo {volume} {83}},\ \bibinfo
  {pages} {034010} (\bibinfo {year} {2011})},\ \Eprint
  {https://arxiv.org/abs/1010.3397} {arXiv:1010.3397 [hep-ph]} \BibitemShut
  {NoStop}%
\bibitem [{\citenamefont {L\"u}\ and\ \citenamefont {Dong}(2016)}]{Lu:2016cwr}%
  \BibitemOpen
  \bibfield  {author} {\bibinfo {author} {\bibfnamefont {Q.-F.}\ \bibnamefont
  {L\"u}}\ and\ \bibinfo {author} {\bibfnamefont {Y.-B.}\ \bibnamefont
  {Dong}},\ }\bibfield  {title} {\bibinfo {title} {{X(4140) , X(4274) , X(4500)
  , and X(4700) in the relativized quark model}},\ }\href
  {https://doi.org/10.1103/PhysRevD.94.074007} {\bibfield  {journal} {\bibinfo
  {journal} {Phys. Rev. D}\ }\textbf {\bibinfo {volume} {94}},\ \bibinfo
  {pages} {074007} (\bibinfo {year} {2016})},\ \Eprint
  {https://arxiv.org/abs/1607.05570} {arXiv:1607.05570 [hep-ph]} \BibitemShut
  {NoStop}%
\bibitem [{\citenamefont {Chen}\ \emph {et~al.}(2013)\citenamefont {Chen},
  \citenamefont {Jin}, \citenamefont {Kleiv}, \citenamefont {Steele},
  \citenamefont {Wang},\ and\ \citenamefont {Xu}}]{Chen:2013pya}%
  \BibitemOpen
  \bibfield  {author} {\bibinfo {author} {\bibfnamefont {W.}~\bibnamefont
  {Chen}}, \bibinfo {author} {\bibfnamefont {H.-y.}\ \bibnamefont {Jin}},
  \bibinfo {author} {\bibfnamefont {R.~T.}\ \bibnamefont {Kleiv}}, \bibinfo
  {author} {\bibfnamefont {T.~G.}\ \bibnamefont {Steele}}, \bibinfo {author}
  {\bibfnamefont {M.}~\bibnamefont {Wang}},\ and\ \bibinfo {author}
  {\bibfnamefont {Q.}~\bibnamefont {Xu}},\ }\bibfield  {title} {\bibinfo
  {title} {{QCD sum-rule interpretation of X(3872) with $J^{PC}=1^{++}$
  mixtures of hybrid charmonium and $\overline{D}D^*$ molecular currents}},\
  }\href {https://doi.org/10.1103/PhysRevD.88.045027} {\bibfield  {journal}
  {\bibinfo  {journal} {Phys. Rev. D}\ }\textbf {\bibinfo {volume} {88}},\
  \bibinfo {pages} {045027} (\bibinfo {year} {2013})},\ \Eprint
  {https://arxiv.org/abs/1305.0244} {arXiv:1305.0244 [hep-ph]} \BibitemShut
  {NoStop}%
\bibitem [{\citenamefont {Aaij}\ \emph {et~al.}(2021)\citenamefont {Aaij} \emph
  {et~al.}}]{LHCb:2020sey}%
  \BibitemOpen
  \bibfield  {author} {\bibinfo {author} {\bibfnamefont {R.}~\bibnamefont
  {Aaij}} \emph {et~al.} (\bibinfo {collaboration} {LHCb}),\ }\bibfield
  {title} {\bibinfo {title} {{Observation of Multiplicity Dependent Prompt
  $\chi_{c1}(3872)$ and $\psi(2S)$ Production in $pp$ Collisions}},\ }\href
  {https://doi.org/10.1103/PhysRevLett.126.092001} {\bibfield  {journal}
  {\bibinfo  {journal} {Phys. Rev. Lett.}\ }\textbf {\bibinfo {volume} {126}},\
  \bibinfo {pages} {092001} (\bibinfo {year} {2021})},\ \Eprint
  {https://arxiv.org/abs/2009.06619} {arXiv:2009.06619 [hep-ex]} \BibitemShut
  {NoStop}%
\bibitem [{\citenamefont {Albuquerque}\ \emph {et~al.}(2011)\citenamefont
  {Albuquerque}, \citenamefont {Nielsen},\ and\ \citenamefont {Rodrigues~da
  Silva}}]{Albuquerque:2011ix}%
  \BibitemOpen
  \bibfield  {author} {\bibinfo {author} {\bibfnamefont {R.~M.}\ \bibnamefont
  {Albuquerque}}, \bibinfo {author} {\bibfnamefont {M.}~\bibnamefont
  {Nielsen}},\ and\ \bibinfo {author} {\bibfnamefont {R.}~\bibnamefont
  {Rodrigues~da Silva}},\ }\bibfield  {title} {\bibinfo {title} {{Exotic
  $1^{--}$ States in QCD Sum Rules}},\ }\href
  {https://doi.org/10.1103/PhysRevD.84.116004} {\bibfield  {journal} {\bibinfo
  {journal} {Phys. Rev. D}\ }\textbf {\bibinfo {volume} {84}},\ \bibinfo
  {pages} {116004} (\bibinfo {year} {2011})},\ \Eprint
  {https://arxiv.org/abs/1110.2113} {arXiv:1110.2113 [hep-ph]} \BibitemShut
  {NoStop}%
\bibitem [{\citenamefont {Finazzo}\ \emph {et~al.}(2011)\citenamefont
  {Finazzo}, \citenamefont {Nielsen},\ and\ \citenamefont
  {Liu}}]{Finazzo:2011he}%
  \BibitemOpen
  \bibfield  {author} {\bibinfo {author} {\bibfnamefont {S.~I.}\ \bibnamefont
  {Finazzo}}, \bibinfo {author} {\bibfnamefont {M.}~\bibnamefont {Nielsen}},\
  and\ \bibinfo {author} {\bibfnamefont {X.}~\bibnamefont {Liu}},\ }\bibfield
  {title} {\bibinfo {title} {{QCD sum rule calculation for the charmonium-like
  structures in the $J/\psi \phi$ and $J/\psi \omega$ invariant mass
  spectra}},\ }\href {https://doi.org/10.1016/j.physletb.2011.05.042}
  {\bibfield  {journal} {\bibinfo  {journal} {Phys. Lett. B}\ }\textbf
  {\bibinfo {volume} {701}},\ \bibinfo {pages} {101} (\bibinfo {year}
  {2011})},\ \Eprint {https://arxiv.org/abs/1102.2347} {arXiv:1102.2347
  [hep-ph]} \BibitemShut {NoStop}%
\bibitem [{\citenamefont {Close}\ and\ \citenamefont
  {Page}(1996)}]{Close:1995eu}%
  \BibitemOpen
  \bibfield  {author} {\bibinfo {author} {\bibfnamefont {F.~E.}\ \bibnamefont
  {Close}}\ and\ \bibinfo {author} {\bibfnamefont {P.~R.}\ \bibnamefont
  {Page}},\ }\bibfield  {title} {\bibinfo {title} {{Do psi (4040), psi (4160)
  signal hybrid charmonium?}},\ }\href
  {https://doi.org/10.1016/0370-2693(95)01265-6} {\bibfield  {journal}
  {\bibinfo  {journal} {Phys. Lett. B}\ }\textbf {\bibinfo {volume} {366}},\
  \bibinfo {pages} {323} (\bibinfo {year} {1996})},\ \Eprint
  {https://arxiv.org/abs/hep-ph/9507407} {arXiv:hep-ph/9507407} \BibitemShut
  {NoStop}%
\bibitem [{\citenamefont {Suzuki}(2005)}]{Suzuki:2005ha}%
  \BibitemOpen
  \bibfield  {author} {\bibinfo {author} {\bibfnamefont {M.}~\bibnamefont
  {Suzuki}},\ }\bibfield  {title} {\bibinfo {title} {{The X(3872) boson:
  Molecule or charmonium}},\ }\href
  {https://doi.org/10.1103/PhysRevD.72.114013} {\bibfield  {journal} {\bibinfo
  {journal} {Phys. Rev. D}\ }\textbf {\bibinfo {volume} {72}},\ \bibinfo
  {pages} {114013} (\bibinfo {year} {2005})},\ \Eprint
  {https://arxiv.org/abs/hep-ph/0508258} {arXiv:hep-ph/0508258} \BibitemShut
  {NoStop}%
\bibitem [{\citenamefont {Matheus}\ \emph {et~al.}(2007)\citenamefont
  {Matheus}, \citenamefont {Narison}, \citenamefont {Nielsen},\ and\
  \citenamefont {Richard}}]{Matheus:2006xi}%
  \BibitemOpen
  \bibfield  {author} {\bibinfo {author} {\bibfnamefont {R.~D.}\ \bibnamefont
  {Matheus}}, \bibinfo {author} {\bibfnamefont {S.}~\bibnamefont {Narison}},
  \bibinfo {author} {\bibfnamefont {M.}~\bibnamefont {Nielsen}},\ and\ \bibinfo
  {author} {\bibfnamefont {J.~M.}\ \bibnamefont {Richard}},\ }\bibfield
  {title} {\bibinfo {title} {{Can the X(3872) be a 1++ four-quark state?}},\
  }\href {https://doi.org/10.1103/PhysRevD.75.014005} {\bibfield  {journal}
  {\bibinfo  {journal} {Phys. Rev. D}\ }\textbf {\bibinfo {volume} {75}},\
  \bibinfo {pages} {014005} (\bibinfo {year} {2007})},\ \Eprint
  {https://arxiv.org/abs/hep-ph/0608297} {arXiv:hep-ph/0608297} \BibitemShut
  {NoStop}%
\bibitem [{\citenamefont {Thomas}\ and\ \citenamefont
  {Close}(2008)}]{Thomas:2008ja}%
  \BibitemOpen
  \bibfield  {author} {\bibinfo {author} {\bibfnamefont {C.~E.}\ \bibnamefont
  {Thomas}}\ and\ \bibinfo {author} {\bibfnamefont {F.~E.}\ \bibnamefont
  {Close}},\ }\bibfield  {title} {\bibinfo {title} {{Is X(3872) a molecule?}},\
  }\href {https://doi.org/10.1103/PhysRevD.78.034007} {\bibfield  {journal}
  {\bibinfo  {journal} {Phys. Rev. D}\ }\textbf {\bibinfo {volume} {78}},\
  \bibinfo {pages} {034007} (\bibinfo {year} {2008})},\ \Eprint
  {https://arxiv.org/abs/0805.3653} {arXiv:0805.3653 [hep-ph]} \BibitemShut
  {NoStop}%
\bibitem [{\citenamefont {Liu}\ \emph {et~al.}(2008)\citenamefont {Liu},
  \citenamefont {Liu}, \citenamefont {Deng},\ and\ \citenamefont
  {Zhu}}]{Liu:2008fh}%
  \BibitemOpen
  \bibfield  {author} {\bibinfo {author} {\bibfnamefont {Y.-R.}\ \bibnamefont
  {Liu}}, \bibinfo {author} {\bibfnamefont {X.}~\bibnamefont {Liu}}, \bibinfo
  {author} {\bibfnamefont {W.-Z.}\ \bibnamefont {Deng}},\ and\ \bibinfo
  {author} {\bibfnamefont {S.-L.}\ \bibnamefont {Zhu}},\ }\bibfield  {title}
  {\bibinfo {title} {{Is $X(3872) $ Really a Molecular State?}},\ }\href
  {https://doi.org/10.1140/epjc/s10052-008-0640-4} {\bibfield  {journal}
  {\bibinfo  {journal} {Eur. Phys. J. C}\ }\textbf {\bibinfo {volume} {56}},\
  \bibinfo {pages} {63} (\bibinfo {year} {2008})},\ \Eprint
  {https://arxiv.org/abs/0801.3540} {arXiv:0801.3540 [hep-ph]} \BibitemShut
  {NoStop}%
\bibitem [{\citenamefont {Chen}\ \emph {et~al.}(2019)\citenamefont {Chen},
  \citenamefont {Zhang}, \citenamefont {Huang}, \citenamefont {Steele},\ and\
  \citenamefont {Jin}}]{Chen:2019buw}%
  \BibitemOpen
  \bibfield  {author} {\bibinfo {author} {\bibfnamefont {Z.-S.}\ \bibnamefont
  {Chen}}, \bibinfo {author} {\bibfnamefont {Z.-F.}\ \bibnamefont {Zhang}},
  \bibinfo {author} {\bibfnamefont {Z.-R.}\ \bibnamefont {Huang}}, \bibinfo
  {author} {\bibfnamefont {T.~G.}\ \bibnamefont {Steele}},\ and\ \bibinfo
  {author} {\bibfnamefont {H.-Y.}\ \bibnamefont {Jin}},\ }\bibfield  {title}
  {\bibinfo {title} {{Vector and scalar mesons\textquoteright{} mixing from QCD
  sum rules}},\ }\href {https://doi.org/10.1007/JHEP12(2019)066} {\bibfield
  {journal} {\bibinfo  {journal} {JHEP}\ }\textbf {\bibinfo {volume} {12}},\
  \bibinfo {pages} {066}},\ \Eprint {https://arxiv.org/abs/1903.06381}
  {arXiv:1903.06381 [hep-ph]} \BibitemShut {NoStop}%
\bibitem [{\citenamefont {Harnett}\ \emph {et~al.}(2011)\citenamefont
  {Harnett}, \citenamefont {Kleiv}, \citenamefont {Moats},\ and\ \citenamefont
  {Steele}}]{Harnett:2008cw}%
  \BibitemOpen
  \bibfield  {author} {\bibinfo {author} {\bibfnamefont {D.}~\bibnamefont
  {Harnett}}, \bibinfo {author} {\bibfnamefont {R.~T.}\ \bibnamefont {Kleiv}},
  \bibinfo {author} {\bibfnamefont {K.}~\bibnamefont {Moats}},\ and\ \bibinfo
  {author} {\bibfnamefont {T.~G.}\ \bibnamefont {Steele}},\ }\bibfield  {title}
  {\bibinfo {title} {{Near-Maximal Mixing of Scalar Gluonium and Quark Mesons:
  A Gaussian Sum-Rule Analysis}},\ }\href
  {https://doi.org/10.1016/j.nuclphysa.2010.12.005} {\bibfield  {journal}
  {\bibinfo  {journal} {Nucl. Phys. A}\ }\textbf {\bibinfo {volume} {850}},\
  \bibinfo {pages} {110} (\bibinfo {year} {2011})},\ \Eprint
  {https://arxiv.org/abs/0804.2195} {arXiv:0804.2195 [hep-ph]} \BibitemShut
  {NoStop}%
\bibitem [{\citenamefont {Palameta}\ \emph
  {et~al.}(2018{\natexlab{a}})\citenamefont {Palameta}, \citenamefont {Ho},
  \citenamefont {Harnett},\ and\ \citenamefont {Steele}}]{Palameta:2017ols}%
  \BibitemOpen
  \bibfield  {author} {\bibinfo {author} {\bibfnamefont {A.}~\bibnamefont
  {Palameta}}, \bibinfo {author} {\bibfnamefont {J.}~\bibnamefont {Ho}},
  \bibinfo {author} {\bibfnamefont {D.}~\bibnamefont {Harnett}},\ and\ \bibinfo
  {author} {\bibfnamefont {T.~G.}\ \bibnamefont {Steele}},\ }\bibfield  {title}
  {\bibinfo {title} {{QCD sum-rules analysis of vector ($1^{--}$) heavy
  quarkonium meson-hybrid mixing}},\ }\href
  {https://doi.org/10.1103/PhysRevD.97.034001} {\bibfield  {journal} {\bibinfo
  {journal} {Phys. Rev. D}\ }\textbf {\bibinfo {volume} {97}},\ \bibinfo
  {pages} {034001} (\bibinfo {year} {2018}{\natexlab{a}})},\ \Eprint
  {https://arxiv.org/abs/1707.00063} {arXiv:1707.00063 [hep-ph]} \BibitemShut
  {NoStop}%
\bibitem [{\citenamefont {Palameta}\ \emph
  {et~al.}(2018{\natexlab{b}})\citenamefont {Palameta}, \citenamefont
  {Harnett},\ and\ \citenamefont {Steele}}]{Palameta:2018yce}%
  \BibitemOpen
  \bibfield  {author} {\bibinfo {author} {\bibfnamefont {A.}~\bibnamefont
  {Palameta}}, \bibinfo {author} {\bibfnamefont {D.}~\bibnamefont {Harnett}},\
  and\ \bibinfo {author} {\bibfnamefont {T.~G.}\ \bibnamefont {Steele}},\
  }\bibfield  {title} {\bibinfo {title} {{Meson-Hybrid Mixing in
  $J^{PC}=1^{++}$ Heavy Quarkonium from QCD Sum-Rules}},\ }\href
  {https://doi.org/10.1103/PhysRevD.98.074014} {\bibfield  {journal} {\bibinfo
  {journal} {Phys. Rev. D}\ }\textbf {\bibinfo {volume} {98}},\ \bibinfo
  {pages} {074014} (\bibinfo {year} {2018}{\natexlab{b}})},\ \Eprint
  {https://arxiv.org/abs/1805.04230} {arXiv:1805.04230 [hep-ph]} \BibitemShut
  {NoStop}%
\bibitem [{\citenamefont {Wang}\ and\ \citenamefont
  {Huang}(2014)}]{Wang:2013daa}%
  \BibitemOpen
  \bibfield  {author} {\bibinfo {author} {\bibfnamefont {Z.-G.}\ \bibnamefont
  {Wang}}\ and\ \bibinfo {author} {\bibfnamefont {T.}~\bibnamefont {Huang}},\
  }\bibfield  {title} {\bibinfo {title} {{Possible assignments of the
  $X(3872)$, $Z_c(3900)$ and $Z_b(10610)$ as axial-vector molecular states}},\
  }\href {https://doi.org/10.1140/epjc/s10052-014-2891-6} {\bibfield  {journal}
  {\bibinfo  {journal} {Eur. Phys. J. C}\ }\textbf {\bibinfo {volume} {74}},\
  \bibinfo {pages} {2891} (\bibinfo {year} {2014})},\ \Eprint
  {https://arxiv.org/abs/1312.7489} {arXiv:1312.7489 [hep-ph]} \BibitemShut
  {NoStop}%
\bibitem [{\citenamefont {Wang}(2014)}]{Wang:2014gwa}%
  \BibitemOpen
  \bibfield  {author} {\bibinfo {author} {\bibfnamefont {Z.-G.}\ \bibnamefont
  {Wang}},\ }\bibfield  {title} {\bibinfo {title} {{Reanalysis of the
  $Y(3940)$, $Y(4140)$, $Z_c(4020)$, $Z_c(4025)$ and $Z_b(10650)$ as molecular
  states with QCD sum rules}},\ }\href
  {https://doi.org/10.1140/epjc/s10052-014-2963-7} {\bibfield  {journal}
  {\bibinfo  {journal} {Eur. Phys. J. C}\ }\textbf {\bibinfo {volume} {74}},\
  \bibinfo {pages} {2963} (\bibinfo {year} {2014})},\ \Eprint
  {https://arxiv.org/abs/1403.0810} {arXiv:1403.0810 [hep-ph]} \BibitemShut
  {NoStop}%
\bibitem [{\citenamefont {Gong}\ \emph {et~al.}(2016)\citenamefont {Gong},
  \citenamefont {Guo}, \citenamefont {Meng}, \citenamefont {Tang},
  \citenamefont {Wang},\ and\ \citenamefont {Zheng}}]{Gong:2016hlt}%
  \BibitemOpen
  \bibfield  {author} {\bibinfo {author} {\bibfnamefont {Q.-R.}\ \bibnamefont
  {Gong}}, \bibinfo {author} {\bibfnamefont {Z.-H.}\ \bibnamefont {Guo}},
  \bibinfo {author} {\bibfnamefont {C.}~\bibnamefont {Meng}}, \bibinfo {author}
  {\bibfnamefont {G.-Y.}\ \bibnamefont {Tang}}, \bibinfo {author}
  {\bibfnamefont {Y.-F.}\ \bibnamefont {Wang}},\ and\ \bibinfo {author}
  {\bibfnamefont {H.-Q.}\ \bibnamefont {Zheng}},\ }\bibfield  {title} {\bibinfo
  {title} {{$Z_c(3900)$ as a $D\bar{D}^*$ molecule from the pole counting
  rule}},\ }\href {https://doi.org/10.1103/PhysRevD.94.114019} {\bibfield
  {journal} {\bibinfo  {journal} {Phys. Rev. D}\ }\textbf {\bibinfo {volume}
  {94}},\ \bibinfo {pages} {114019} (\bibinfo {year} {2016})},\ \Eprint
  {https://arxiv.org/abs/1604.08836} {arXiv:1604.08836 [hep-ph]} \BibitemShut
  {NoStop}%
\bibitem [{\citenamefont {Albuquerque}\ and\ \citenamefont
  {Matheus}(2015)}]{Albuquerque:2015kia}%
  \BibitemOpen
  \bibfield  {author} {\bibinfo {author} {\bibfnamefont {R.~M.}\ \bibnamefont
  {Albuquerque}}\ and\ \bibinfo {author} {\bibfnamefont {R.~D.}\ \bibnamefont
  {Matheus}},\ }\bibfield  {title} {\bibinfo {title} {{The $J/\Psi\Phi$ decay
  channel of the Y(4140) molecular state}},\ }\href
  {https://doi.org/10.1016/j.nuclphysbps.2015.01.032} {\bibfield  {journal}
  {\bibinfo  {journal} {Nucl. Part. Phys. Proc.}\ }\textbf {\bibinfo {volume}
  {258-259}},\ \bibinfo {pages} {148} (\bibinfo {year} {2015})}\BibitemShut
  {NoStop}%
\bibitem [{\citenamefont {Cui}\ \emph {et~al.}(2013)\citenamefont {Cui},
  \citenamefont {Liu},\ and\ \citenamefont {Huang}}]{Cui:2013xla}%
  \BibitemOpen
  \bibfield  {author} {\bibinfo {author} {\bibfnamefont {C.-Y.}\ \bibnamefont
  {Cui}}, \bibinfo {author} {\bibfnamefont {Y.-L.}\ \bibnamefont {Liu}},\ and\
  \bibinfo {author} {\bibfnamefont {M.-Q.}\ \bibnamefont {Huang}},\ }\bibfield
  {title} {\bibinfo {title} {{Could $Z_c$(4025) be a $J^P$ = $1^+ D^*
  \bar{D^*}$ molecular state?}},\ }\href
  {https://doi.org/10.1140/epjc/s10052-013-2661-x} {\bibfield  {journal}
  {\bibinfo  {journal} {Eur. Phys. J. C}\ }\textbf {\bibinfo {volume} {73}},\
  \bibinfo {pages} {2661} (\bibinfo {year} {2013})},\ \Eprint
  {https://arxiv.org/abs/1308.3625} {arXiv:1308.3625 [hep-ph]} \BibitemShut
  {NoStop}%
\bibitem [{\citenamefont {Chen}\ \emph {et~al.}(2014)\citenamefont {Chen},
  \citenamefont {Steele}, \citenamefont {Du},\ and\ \citenamefont
  {Zhu}}]{Chen:2013omd}%
  \BibitemOpen
  \bibfield  {author} {\bibinfo {author} {\bibfnamefont {W.}~\bibnamefont
  {Chen}}, \bibinfo {author} {\bibfnamefont {T.~G.}\ \bibnamefont {Steele}},
  \bibinfo {author} {\bibfnamefont {M.-L.}\ \bibnamefont {Du}},\ and\ \bibinfo
  {author} {\bibfnamefont {S.-L.}\ \bibnamefont {Zhu}},\ }\bibfield  {title}
  {\bibinfo {title} {{$D^*\bar D^*$ molecule interpretation of $Z_c(4025)$}},\
  }\href {https://doi.org/10.1140/epjc/s10052-014-2773-y} {\bibfield  {journal}
  {\bibinfo  {journal} {Eur. Phys. J. C}\ }\textbf {\bibinfo {volume} {74}},\
  \bibinfo {pages} {2773} (\bibinfo {year} {2014})},\ \Eprint
  {https://arxiv.org/abs/1308.5060} {arXiv:1308.5060 [hep-ph]} \BibitemShut
  {NoStop}%
\bibitem [{\citenamefont {Chen}\ \emph {et~al.}(2015)\citenamefont {Chen},
  \citenamefont {Steele}, \citenamefont {Chen},\ and\ \citenamefont
  {Zhu}}]{Chen:2015ata}%
  \BibitemOpen
  \bibfield  {author} {\bibinfo {author} {\bibfnamefont {W.}~\bibnamefont
  {Chen}}, \bibinfo {author} {\bibfnamefont {T.~G.}\ \bibnamefont {Steele}},
  \bibinfo {author} {\bibfnamefont {H.-X.}\ \bibnamefont {Chen}},\ and\
  \bibinfo {author} {\bibfnamefont {S.-L.}\ \bibnamefont {Zhu}},\ }\bibfield
  {title} {\bibinfo {title} {{Mass spectra of Zc and Zb exotic states as hadron
  molecules}},\ }\href {https://doi.org/10.1103/PhysRevD.92.054002} {\bibfield
  {journal} {\bibinfo  {journal} {Phys. Rev. D}\ }\textbf {\bibinfo {volume}
  {92}},\ \bibinfo {pages} {054002} (\bibinfo {year} {2015})},\ \Eprint
  {https://arxiv.org/abs/1505.05619} {arXiv:1505.05619 [hep-ph]} \BibitemShut
  {NoStop}%
\bibitem [{\citenamefont {Liu}\ \emph {et~al.}(2009)\citenamefont {Liu},
  \citenamefont {Luo}, \citenamefont {Liu},\ and\ \citenamefont
  {Zhu}}]{Liu:2008tn}%
  \BibitemOpen
  \bibfield  {author} {\bibinfo {author} {\bibfnamefont {X.}~\bibnamefont
  {Liu}}, \bibinfo {author} {\bibfnamefont {Z.-G.}\ \bibnamefont {Luo}},
  \bibinfo {author} {\bibfnamefont {Y.-R.}\ \bibnamefont {Liu}},\ and\ \bibinfo
  {author} {\bibfnamefont {S.-L.}\ \bibnamefont {Zhu}},\ }\bibfield  {title}
  {\bibinfo {title} {{X(3872) and Other Possible Heavy Molecular States}},\
  }\href {https://doi.org/10.1140/epjc/s10052-009-1020-4} {\bibfield  {journal}
  {\bibinfo  {journal} {Eur. Phys. J. C}\ }\textbf {\bibinfo {volume} {61}},\
  \bibinfo {pages} {411} (\bibinfo {year} {2009})},\ \Eprint
  {https://arxiv.org/abs/0808.0073} {arXiv:0808.0073 [hep-ph]} \BibitemShut
  {NoStop}%
\bibitem [{\citenamefont {Dong}\ \emph {et~al.}(2009)\citenamefont {Dong},
  \citenamefont {Faessler}, \citenamefont {Gutsche}, \citenamefont
  {Kovalenko},\ and\ \citenamefont {Lyubovitskij}}]{Dong:2009yp}%
  \BibitemOpen
  \bibfield  {author} {\bibinfo {author} {\bibfnamefont {Y.}~\bibnamefont
  {Dong}}, \bibinfo {author} {\bibfnamefont {A.}~\bibnamefont {Faessler}},
  \bibinfo {author} {\bibfnamefont {T.}~\bibnamefont {Gutsche}}, \bibinfo
  {author} {\bibfnamefont {S.}~\bibnamefont {Kovalenko}},\ and\ \bibinfo
  {author} {\bibfnamefont {V.~E.}\ \bibnamefont {Lyubovitskij}},\ }\bibfield
  {title} {\bibinfo {title} {{X(3872) as a hadronic molecule and its decays to
  charmonium states and pions}},\ }\href
  {https://doi.org/10.1103/PhysRevD.79.094013} {\bibfield  {journal} {\bibinfo
  {journal} {Phys. Rev. D}\ }\textbf {\bibinfo {volume} {79}},\ \bibinfo
  {pages} {094013} (\bibinfo {year} {2009})},\ \Eprint
  {https://arxiv.org/abs/0903.5416} {arXiv:0903.5416 [hep-ph]} \BibitemShut
  {NoStop}%
\bibitem [{\citenamefont {Matheus}\ \emph {et~al.}(2009)\citenamefont
  {Matheus}, \citenamefont {Navarra}, \citenamefont {Nielsen},\ and\
  \citenamefont {Zanetti}}]{Matheus:2009vq}%
  \BibitemOpen
  \bibfield  {author} {\bibinfo {author} {\bibfnamefont {R.~D.}\ \bibnamefont
  {Matheus}}, \bibinfo {author} {\bibfnamefont {F.~S.}\ \bibnamefont
  {Navarra}}, \bibinfo {author} {\bibfnamefont {M.}~\bibnamefont {Nielsen}},\
  and\ \bibinfo {author} {\bibfnamefont {C.~M.}\ \bibnamefont {Zanetti}},\
  }\bibfield  {title} {\bibinfo {title} {{QCD Sum Rules for the X(3872) as a
  mixed molecule-charmoniun state}},\ }\href
  {https://doi.org/10.1103/PhysRevD.80.056002} {\bibfield  {journal} {\bibinfo
  {journal} {Phys. Rev. D}\ }\textbf {\bibinfo {volume} {80}},\ \bibinfo
  {pages} {056002} (\bibinfo {year} {2009})},\ \Eprint
  {https://arxiv.org/abs/0907.2683} {arXiv:0907.2683 [hep-ph]} \BibitemShut
  {NoStop}%
\bibitem [{\citenamefont {Jackson}(1979)}]{Jackson:1979eu}%
  \BibitemOpen
  \bibfield  {author} {\bibinfo {author} {\bibfnamefont {J.~D.}\ \bibnamefont
  {Jackson}},\ }\bibfield  {title} {\bibinfo {title} {{Charmonium Chi States:
  Radiative and Hadronic Widths, Spin of the Gluon}},\ }\href
  {https://doi.org/10.1016/0370-2693(79)90030-3} {\bibfield  {journal}
  {\bibinfo  {journal} {Phys. Lett. B}\ }\textbf {\bibinfo {volume} {87}},\
  \bibinfo {pages} {106} (\bibinfo {year} {1979})}\BibitemShut {NoStop}%
\bibitem [{\citenamefont {Wang}(2017)}]{Wang:2016wwe}%
  \BibitemOpen
  \bibfield  {author} {\bibinfo {author} {\bibfnamefont {Z.-G.}\ \bibnamefont
  {Wang}},\ }\bibfield  {title} {\bibinfo {title} {{Analysis of the $Y(4220)$
  and $Y(4390)$ as molecular states with QCD sum rules}},\ }\href
  {https://doi.org/10.1088/1674-1137/41/8/083103} {\bibfield  {journal}
  {\bibinfo  {journal} {Chin. Phys. C}\ }\textbf {\bibinfo {volume} {41}},\
  \bibinfo {pages} {083103} (\bibinfo {year} {2017})},\ \Eprint
  {https://arxiv.org/abs/1611.03250} {arXiv:1611.03250 [hep-ph]} \BibitemShut
  {NoStop}%
\bibitem [{\citenamefont {Wang}(2018)}]{Wang:2018rfw}%
  \BibitemOpen
  \bibfield  {author} {\bibinfo {author} {\bibfnamefont {Z.-G.}\ \bibnamefont
  {Wang}},\ }\bibfield  {title} {\bibinfo {title} {{Vector tetraquark state
  candidates: $Y(4260/4220)$, $Y(4360/4320)$, $Y(4390)$ and $Y(4660/4630)$}},\
  }\href {https://doi.org/10.1140/epjc/s10052-018-5996-5} {\bibfield  {journal}
  {\bibinfo  {journal} {Eur. Phys. J. C}\ }\textbf {\bibinfo {volume} {78}},\
  \bibinfo {pages} {518} (\bibinfo {year} {2018})},\ \Eprint
  {https://arxiv.org/abs/1803.05749} {arXiv:1803.05749 [hep-ph]} \BibitemShut
  {NoStop}%
\bibitem [{\citenamefont {Shifman}\ \emph
  {et~al.}(1979{\natexlab{a}})\citenamefont {Shifman}, \citenamefont
  {Vainshtein},\ and\ \citenamefont {Zakharov}}]{Shifman:1978bx}%
  \BibitemOpen
  \bibfield  {author} {\bibinfo {author} {\bibfnamefont {M.~A.}\ \bibnamefont
  {Shifman}}, \bibinfo {author} {\bibfnamefont {A.~I.}\ \bibnamefont
  {Vainshtein}},\ and\ \bibinfo {author} {\bibfnamefont {V.~I.}\ \bibnamefont
  {Zakharov}},\ }\bibfield  {title} {\bibinfo {title} {{QCD and Resonance
  Physics. Theoretical Foundations}},\ }\href
  {https://doi.org/10.1016/0550-3213(79)90022-1} {\bibfield  {journal}
  {\bibinfo  {journal} {Nucl. Phys. B}\ }\textbf {\bibinfo {volume} {147}},\
  \bibinfo {pages} {385} (\bibinfo {year} {1979}{\natexlab{a}})}\BibitemShut
  {NoStop}%
\bibitem [{\citenamefont {Shifman}\ \emph
  {et~al.}(1979{\natexlab{b}})\citenamefont {Shifman}, \citenamefont
  {Vainshtein},\ and\ \citenamefont {Zakharov}}]{Shifman:1978by}%
  \BibitemOpen
  \bibfield  {author} {\bibinfo {author} {\bibfnamefont {M.~A.}\ \bibnamefont
  {Shifman}}, \bibinfo {author} {\bibfnamefont {A.~I.}\ \bibnamefont
  {Vainshtein}},\ and\ \bibinfo {author} {\bibfnamefont {V.~I.}\ \bibnamefont
  {Zakharov}},\ }\bibfield  {title} {\bibinfo {title} {{QCD and Resonance
  Physics: Applications}},\ }\href
  {https://doi.org/10.1016/0550-3213(79)90023-3} {\bibfield  {journal}
  {\bibinfo  {journal} {Nucl. Phys. B}\ }\textbf {\bibinfo {volume} {147}},\
  \bibinfo {pages} {448} (\bibinfo {year} {1979}{\natexlab{b}})}\BibitemShut
  {NoStop}%
\bibitem [{\citenamefont {Ho}\ \emph {et~al.}(2018)\citenamefont {Ho},
  \citenamefont {Berg}, \citenamefont {Chen}, \citenamefont {Harnett},\ and\
  \citenamefont {Steele}}]{Ho:2018cat}%
  \BibitemOpen
  \bibfield  {author} {\bibinfo {author} {\bibfnamefont {J.}~\bibnamefont
  {Ho}}, \bibinfo {author} {\bibfnamefont {R.}~\bibnamefont {Berg}}, \bibinfo
  {author} {\bibfnamefont {W.}~\bibnamefont {Chen}}, \bibinfo {author}
  {\bibfnamefont {D.}~\bibnamefont {Harnett}},\ and\ \bibinfo {author}
  {\bibfnamefont {T.~G.}\ \bibnamefont {Steele}},\ }\bibfield  {title}
  {\bibinfo {title} {{Mass Calculations of Light Quarkonium, Exotic
  $J^{PC}=0^{+-}$ Hybrid Mesons from Gaussian Sum-Rules}},\ }\href
  {https://doi.org/10.1103/PhysRevD.98.096020} {\bibfield  {journal} {\bibinfo
  {journal} {Phys. Rev. D}\ }\textbf {\bibinfo {volume} {98}},\ \bibinfo
  {pages} {096020} (\bibinfo {year} {2018})},\ \Eprint
  {https://arxiv.org/abs/1806.02465} {arXiv:1806.02465 [hep-ph]} \BibitemShut
  {NoStop}%
\bibitem [{\citenamefont {Narison}(2009)}]{Narison:2009vj}%
  \BibitemOpen
  \bibfield  {author} {\bibinfo {author} {\bibfnamefont {S.}~\bibnamefont
  {Narison}},\ }\bibfield  {title} {\bibinfo {title} {{1-+ light exotic mesons
  in QCD}},\ }\href {https://doi.org/10.1016/j.physletb.2009.04.012} {\bibfield
   {journal} {\bibinfo  {journal} {Phys. Lett. B}\ }\textbf {\bibinfo {volume}
  {675}},\ \bibinfo {pages} {319} (\bibinfo {year} {2009})},\ \Eprint
  {https://arxiv.org/abs/0903.2266} {arXiv:0903.2266 [hep-ph]} \BibitemShut
  {NoStop}%
\bibitem [{\citenamefont {Mo}\ \emph {et~al.}(2014)\citenamefont {Mo},
  \citenamefont {Cui}, \citenamefont {Liu},\ and\ \citenamefont
  {Huang}}]{Mo:2014nua}%
  \BibitemOpen
  \bibfield  {author} {\bibinfo {author} {\bibfnamefont {Z.}~\bibnamefont
  {Mo}}, \bibinfo {author} {\bibfnamefont {C.-Y.}\ \bibnamefont {Cui}},
  \bibinfo {author} {\bibfnamefont {Y.-L.}\ \bibnamefont {Liu}},\ and\ \bibinfo
  {author} {\bibfnamefont {M.-Q.}\ \bibnamefont {Huang}},\ }\bibfield  {title}
  {\bibinfo {title} {{QCD sum rules study of X(4350)}},\ }\href
  {https://doi.org/10.1088/0253-6102/61/4/16} {\bibfield  {journal} {\bibinfo
  {journal} {Commun. Theor. Phys.}\ }\textbf {\bibinfo {volume} {61}},\
  \bibinfo {pages} {501} (\bibinfo {year} {2014})},\ \Eprint
  {https://arxiv.org/abs/1403.6906} {arXiv:1403.6906 [hep-ph]} \BibitemShut
  {NoStop}%
\bibitem [{\citenamefont {Matheus}\ \emph {et~al.}(2010)\citenamefont
  {Matheus}, \citenamefont {Navarra}, \citenamefont {Nielsen},\ and\
  \citenamefont {Zanetti}}]{Matheus:2010qxa}%
  \BibitemOpen
  \bibfield  {author} {\bibinfo {author} {\bibfnamefont {R.~D.}\ \bibnamefont
  {Matheus}}, \bibinfo {author} {\bibfnamefont {F.~S.}\ \bibnamefont
  {Navarra}}, \bibinfo {author} {\bibfnamefont {M.}~\bibnamefont {Nielsen}},\
  and\ \bibinfo {author} {\bibfnamefont {C.~M.}\ \bibnamefont {Zanetti}},\
  }\bibfield  {title} {\bibinfo {title} {{Understanding the X(3872) with QCD
  sum rules}},\ }\href {https://doi.org/10.1051/epjconf/20100303025} {\bibfield
   {journal} {\bibinfo  {journal} {EPJ Web Conf.}\ }\textbf {\bibinfo {volume}
  {3}},\ \bibinfo {pages} {03025} (\bibinfo {year} {2010})}\BibitemShut
  {NoStop}%
\bibitem [{\citenamefont {Hart}\ \emph {et~al.}(2006)\citenamefont {Hart},
  \citenamefont {McNeile}, \citenamefont {Michael},\ and\ \citenamefont
  {Pickavance}}]{Hart:2006ps}%
  \BibitemOpen
  \bibfield  {author} {\bibinfo {author} {\bibfnamefont {A.}~\bibnamefont
  {Hart}}, \bibinfo {author} {\bibfnamefont {C.}~\bibnamefont {McNeile}},
  \bibinfo {author} {\bibfnamefont {C.}~\bibnamefont {Michael}},\ and\ \bibinfo
  {author} {\bibfnamefont {J.}~\bibnamefont {Pickavance}} (\bibinfo
  {collaboration} {UKQCD}),\ }\bibfield  {title} {\bibinfo {title} {{A Lattice
  study of the masses of singlet 0++ mesons}},\ }\href
  {https://doi.org/10.1103/PhysRevD.74.114504} {\bibfield  {journal} {\bibinfo
  {journal} {Phys. Rev. D}\ }\textbf {\bibinfo {volume} {74}},\ \bibinfo
  {pages} {114504} (\bibinfo {year} {2006})},\ \Eprint
  {https://arxiv.org/abs/hep-lat/0608026} {arXiv:hep-lat/0608026} \BibitemShut
  {NoStop}%
\bibitem [{\citenamefont {Narison}\ \emph {et~al.}(1984)\citenamefont
  {Narison}, \citenamefont {Pak},\ and\ \citenamefont
  {Paver}}]{Narison:1984bv}%
  \BibitemOpen
  \bibfield  {author} {\bibinfo {author} {\bibfnamefont {S.}~\bibnamefont
  {Narison}}, \bibinfo {author} {\bibfnamefont {N.}~\bibnamefont {Pak}},\ and\
  \bibinfo {author} {\bibfnamefont {N.}~\bibnamefont {Paver}},\ }\bibfield
  {title} {\bibinfo {title} {{Meson - Gluonium Mixing From \{QCD\} Sum
  Rules}},\ }\href {https://doi.org/10.1016/0370-2693(84)90613-0} {\ ,\
  \bibinfo {pages} {77} (\bibinfo {year} {1984})}\BibitemShut {NoStop}%
\bibitem [{\citenamefont {Reinders}\ \emph {et~al.}(1985)\citenamefont
  {Reinders}, \citenamefont {Rubinstein},\ and\ \citenamefont
  {Yazaki}}]{Reinders:1984sr}%
  \BibitemOpen
  \bibfield  {author} {\bibinfo {author} {\bibfnamefont {L.~J.}\ \bibnamefont
  {Reinders}}, \bibinfo {author} {\bibfnamefont {H.}~\bibnamefont
  {Rubinstein}},\ and\ \bibinfo {author} {\bibfnamefont {S.}~\bibnamefont
  {Yazaki}},\ }\bibfield  {title} {\bibinfo {title} {{Hadron Properties from
  QCD Sum Rules}},\ }\href {https://doi.org/10.1016/0370-1573(85)90065-1}
  {\bibfield  {journal} {\bibinfo  {journal} {Phys. Rept.}\ }\textbf {\bibinfo
  {volume} {127}},\ \bibinfo {pages} {1} (\bibinfo {year} {1985})}\BibitemShut
  {NoStop}%
\bibitem [{\citenamefont {Narison}(2010)}]{Narison:2010wb}%
  \BibitemOpen
  \bibfield  {author} {\bibinfo {author} {\bibfnamefont {S.}~\bibnamefont
  {Narison}},\ }\bibfield  {title} {\bibinfo {title} {{SVZ sum rules : 30 + 1
  years later}},\ }\href {https://doi.org/10.1016/j.nuclphysbps.2010.10.078}
  {\bibfield  {journal} {\bibinfo  {journal} {Nucl. Phys. B Proc. Suppl.}\
  }\textbf {\bibinfo {volume} {207-208}},\ \bibinfo {pages} {315} (\bibinfo
  {year} {2010})},\ \Eprint {https://arxiv.org/abs/1010.1959} {arXiv:1010.1959
  [hep-ph]} \BibitemShut {NoStop}%
\bibitem [{\citenamefont {Zhang}\ and\ \citenamefont
  {Huang}(2010)}]{Zhang:2009em}%
  \BibitemOpen
  \bibfield  {author} {\bibinfo {author} {\bibfnamefont {J.-R.}\ \bibnamefont
  {Zhang}}\ and\ \bibinfo {author} {\bibfnamefont {M.-Q.}\ \bibnamefont
  {Huang}},\ }\bibfield  {title} {\bibinfo {title} {{{Q
  anti-s}{anti-Q-(prime)s} molecular states in QCD sum rules}},\ }\href
  {https://doi.org/10.1088/0253-6102/54/6/22} {\bibfield  {journal} {\bibinfo
  {journal} {Commun. Theor. Phys.}\ }\textbf {\bibinfo {volume} {54}},\
  \bibinfo {pages} {1075} (\bibinfo {year} {2010})},\ \Eprint
  {https://arxiv.org/abs/0905.4672} {arXiv:0905.4672 [hep-ph]} \BibitemShut
  {NoStop}%
\bibitem [{\citenamefont {Lee}\ \emph {et~al.}(2009)\citenamefont {Lee},
  \citenamefont {Morita},\ and\ \citenamefont {Nielsen}}]{Lee:2008gn}%
  \BibitemOpen
  \bibfield  {author} {\bibinfo {author} {\bibfnamefont {S.~H.}\ \bibnamefont
  {Lee}}, \bibinfo {author} {\bibfnamefont {K.}~\bibnamefont {Morita}},\ and\
  \bibinfo {author} {\bibfnamefont {M.}~\bibnamefont {Nielsen}},\ }\bibfield
  {title} {\bibinfo {title} {{Can the pi+ chi(c1) resonance structures be D*
  anti-D* and D(1) anti-D molecules?}},\ }\href
  {https://doi.org/10.1016/j.nuclphysa.2008.10.012} {\bibfield  {journal}
  {\bibinfo  {journal} {Nucl. Phys. A}\ }\textbf {\bibinfo {volume} {815}},\
  \bibinfo {pages} {29} (\bibinfo {year} {2009})},\ \Eprint
  {https://arxiv.org/abs/0808.0690} {arXiv:0808.0690 [hep-ph]} \BibitemShut
  {NoStop}%
\bibitem [{\citenamefont {Nielsen}\ \emph {et~al.}(2010)\citenamefont
  {Nielsen}, \citenamefont {Navarra},\ and\ \citenamefont
  {Lee}}]{Nielsen:2009uh}%
  \BibitemOpen
  \bibfield  {author} {\bibinfo {author} {\bibfnamefont {M.}~\bibnamefont
  {Nielsen}}, \bibinfo {author} {\bibfnamefont {F.~S.}\ \bibnamefont
  {Navarra}},\ and\ \bibinfo {author} {\bibfnamefont {S.~H.}\ \bibnamefont
  {Lee}},\ }\bibfield  {title} {\bibinfo {title} {{New Charmonium States in QCD
  Sum Rules: A Concise Review}},\ }\href
  {https://doi.org/10.1016/j.physrep.2010.07.005} {\bibfield  {journal}
  {\bibinfo  {journal} {Phys. Rept.}\ }\textbf {\bibinfo {volume} {497}},\
  \bibinfo {pages} {41} (\bibinfo {year} {2010})},\ \Eprint
  {https://arxiv.org/abs/0911.1958} {arXiv:0911.1958 [hep-ph]} \BibitemShut
  {NoStop}%
\bibitem [{\citenamefont {Guo}\ \emph {et~al.}(2018)\citenamefont {Guo},
  \citenamefont {Hanhart}, \citenamefont {Mei\ss{}ner}, \citenamefont {Wang},
  \citenamefont {Zhao},\ and\ \citenamefont {Zou}}]{Guo:2017jvc}%
  \BibitemOpen
  \bibfield  {author} {\bibinfo {author} {\bibfnamefont {F.-K.}\ \bibnamefont
  {Guo}}, \bibinfo {author} {\bibfnamefont {C.}~\bibnamefont {Hanhart}},
  \bibinfo {author} {\bibfnamefont {U.-G.}\ \bibnamefont {Mei\ss{}ner}},
  \bibinfo {author} {\bibfnamefont {Q.}~\bibnamefont {Wang}}, \bibinfo {author}
  {\bibfnamefont {Q.}~\bibnamefont {Zhao}},\ and\ \bibinfo {author}
  {\bibfnamefont {B.-S.}\ \bibnamefont {Zou}},\ }\bibfield  {title} {\bibinfo
  {title} {{Hadronic molecules}},\ }\href
  {https://doi.org/10.1103/RevModPhys.90.015004} {\bibfield  {journal}
  {\bibinfo  {journal} {Rev. Mod. Phys.}\ }\textbf {\bibinfo {volume} {90}},\
  \bibinfo {pages} {015004} (\bibinfo {year} {2018})},\ \Eprint
  {https://arxiv.org/abs/1705.00141} {arXiv:1705.00141 [hep-ph]} \BibitemShut
  {NoStop}%
\end{thebibliography}%

\end{document}